\documentclass[aps, prx, twocolumn, superscriptaddress]{revtex4-2}
\usepackage{hyperref, graphicx, xcolor, amsmath, amssymb}
\usepackage{bm, dcolumn}
\usepackage{braket, dsfont, textcomp, siunitx}
\hypersetup{colorlinks, linkcolor=teal, citecolor=blue, urlcolor=purple}

\newcommand{\EJ}{5.178}
\newcommand{\EC}{0.4144}
\newcommand{\EL}{0.18}
\newcommand{\qubitfreq}{1.8}

\newcommand{\LKB}{Laboratoire Kastler Brossel, Sorbonne Université, CNRS,
ENS-Université PSL, Collège de France, 4 place Jussieu, 75005 Paris, France}
\newcommand{\LPENS}{Laboratoire de Physique de l’École Normale Supérieure,
ENS, Université PSL, CNRS, Sorbonne Université, 75005 Paris, France}
\newcommand{\QUANTRO}{Quantronics group, Université Paris-Saclay, CEA,
CNRS, SPEC, 91191 Gif-sur-Yvette Cedex, France}
\newcommand{\INRIA}{Centre Automatique et Systèmes, Mines Paris, Inria}
\newcommand{\AB}{Alice \& Bob, 53 Bd du Général Martial Valin, 75015 Paris, France}

\begin{document}

\title{High-sensitivity AC-charge detection with a MHz-frequency fluxonium qubit}

\author{B.-L. Najera-Santos}
\affiliation{\LKB}
\author{R. Rousseau}
\affiliation{\LKB}
\affiliation{\AB}
\author{K. Gerashchenko}
\affiliation{\LKB}
\author{H. Patange}
\affiliation{\LKB}
\author{A. Riva}
\affiliation{\LPENS}
\affiliation{\INRIA}
\author{M. Villiers}
\affiliation{\LPENS}
\affiliation{\INRIA}
\author{T. Briant}
\affiliation{\LKB}
\author{P.-F. Cohadon}
\affiliation{\LKB}
\author{A. Heidmann}
\affiliation{\LKB}
\author{J. Palomo}
\affiliation{\LPENS}
\author{M. Rosticher}
\affiliation{\LPENS}
\author{H. le Sueur}
\affiliation{\QUANTRO}
\author{A. Sarlette}
\affiliation{\LPENS}
\affiliation{\INRIA}
\author{W.~C.~Smith}
\altaffiliation{Present address: Google Quantum AI, Santa Barbara, CA}
\affiliation{\LPENS}
\affiliation{\INRIA}
\author{Z. Leghtas}
\affiliation{\LPENS}
\affiliation{\INRIA}
\author{E. Flurin}
\affiliation{\QUANTRO}
\author{T. Jacqmin}
\affiliation{\LKB}
\author{S. Deléglise}
 \email{samuel.deleglise@lkb.upmc.fr}
\affiliation{\LKB}

\date{\today}

\begin{abstract}

Owing to their strong dipole moment and long coherence times, superconducting qubits have demonstrated remarkable success in hybrid quantum circuits. However, most qubit architectures are limited to the GHz frequency range, severely constraining the class of systems they can interact with. 
The fluxonium qubit, on the other hand, can be biased to very low frequency while being manipulated and read out with standard microwave techniques. 
Here, we design and operate a heavy fluxonium with an unprecedentedly low transition frequency of $\SI{\qubitfreq}{\mega\hertz}$. 
We demonstrate resolved sideband cooling of the ``hot'' qubit transition with a final ground state population of $\SI{97.7}{\percent}$, corresponding to an effective temperature of $\SI{23}{\micro\kelvin}$. We further demonstrate coherent manipulation with coherence times $T_1=\SI{34}{\micro\second}$, $T_2^*=\SI{39}{\micro\second}$, and single-shot readout of the qubit state. 
Importantly, by directly addressing the qubit transition with a capacitively coupled waveguide, we showcase its high sensitivity to a radio-frequency field. 
Through cyclic qubit preparation and interrogation, we transform this low-frequency fluxonium qubit into a frequency-resolved charge sensor. 
This method results in a charge sensitivity of $33~\mu\mathrm{e}/\sqrt{\mathrm{Hz}}$, or an energy sensitivity (in joules per hertz) of 2.8~$\hbar$. This method rivals state-of-the-art transport-based devices, while maintaining inherent insensitivity to DC charge noise. 
The high charge sensitivity combined with large capacitive shunt unlocks new avenues for exploring quantum phenomena in the 1--$\SI{10}{\mega\hertz}$ range, such as the strong-coupling regime with a resonant macroscopic mechanical resonator.
\end{abstract}

\maketitle

\begin{figure}[t]
    \includegraphics[width=0.46\textwidth]{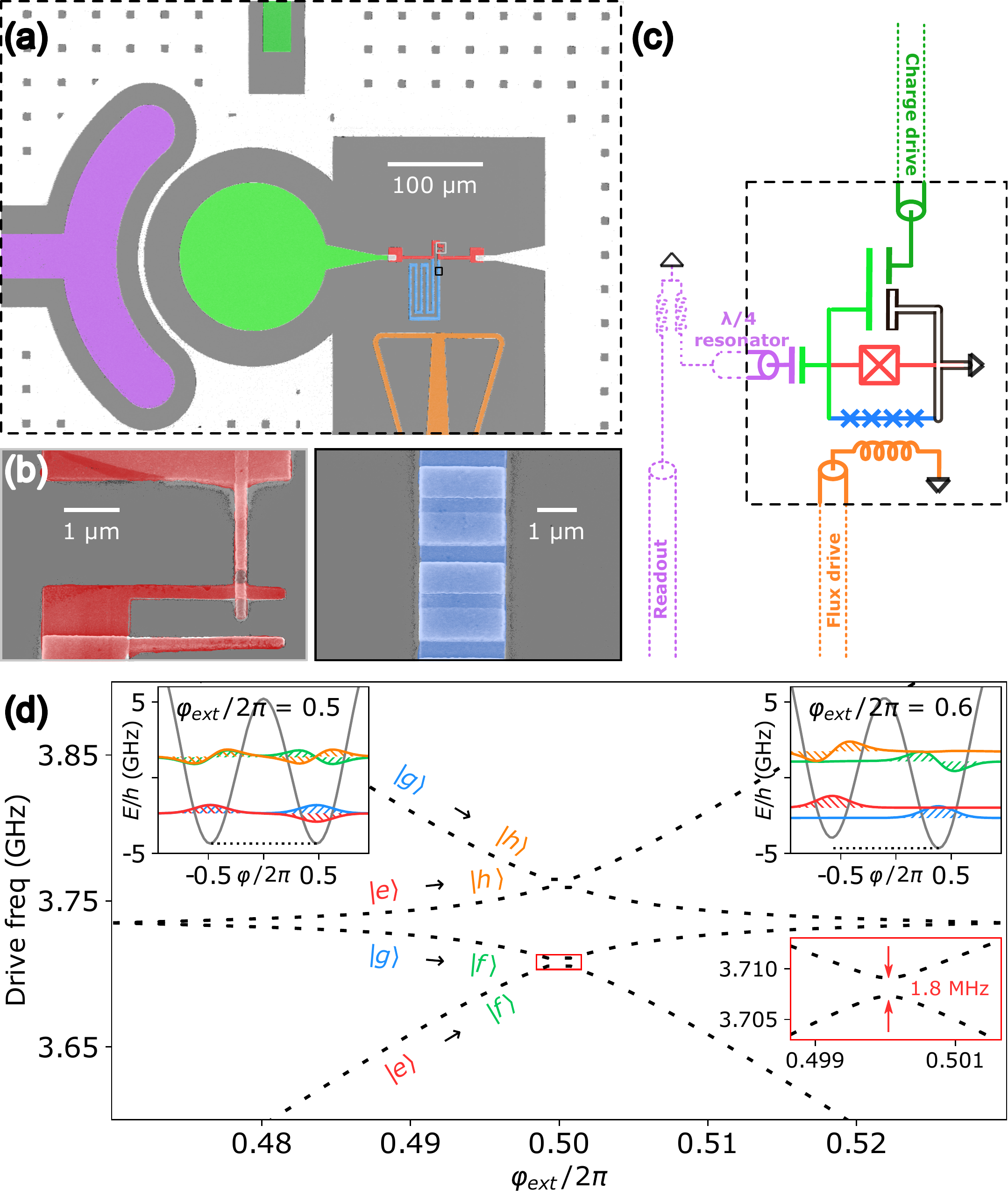}
    \caption{Circuit diagram and implementation of the fluxonium qubit, controls and readout. Optical micrograph (a) and circuit diagram (c) of the fluxonium qubit composed of a capacitor (green), an array of 360 Josephson junctions (blue) and a single junction (red). The qubit is capacitively coupled to the readout resonator (purple). The magnetic flux through the superconducting loop can be rapidly tuned via the current passing through the flux line (orange) and the qubit is capacitively coupled to a charge line (dark green). (b) Scanning electron micrograph of the fluxonium single junction (red) and four junctions of the array (blue). (d) Two tone spectrum centered on the flux frustration point $\varphi_\mathrm{ext}/2\pi = 0.5$. The colorscale in the image represents the phase of the reflected probe pulse. The fit to the data (dotted lines) yields the qubit parameters $E_J/h = \EJ$ GHz, $E_C/h = \EC$ GHz, and $E_L/h=\EL$ GHz. Central inset: detailed view of the avoided-crossing near the frustration point. The left and right insets represent the energy-level diagrams of the 4 lowest eigenstates, with the potential in grey and the wave-functions in colors for an external flux of $\varphi_\mathrm{ext}/2\pi = 0.5$ and $\varphi_\mathrm{ext}/2\pi = 0.6$ respectively. }
    \label{fig:figure1}
\end{figure}

\section{Introduction}

Superconducting qubits consist of engineered quantum systems with lowest-level spacings designed to host a two-level system which can be manipulated and read-out via its dipolar interaction with electromagnetic fields. 
Their strong dipole moment is also beneficial to interface them with other physical systems. 
For instance, fluorescence from individual electronic spins was successfully detected using a superconducting qubit-based microwave-photon detector~\cite{Wang2023} operating close to 7~GHz. 
Additionally, in the realm of circuit quantum acousto-dynamics (cQAD), the coupling between a qubit and a piezoelectric resonator is used to detect and manipulate the phononic state, typically within the 2-10 GHz range~\cite{Oconnell2010, Chu2018, Satzinger2018, Arrangoiz2019}. 
However, adapting these sensing schemes to lower frequencies, below the conventional operating frequency of superconducting qubits, introduces distinct challenges.

First, superconducting qubits are read out thanks to the dispersive shift imparted to a nearby superconducting resonator. 
As the dispersive shift quickly drops for a cavity detuning exceeding the qubit anharmonicity, weakly anharmonic qubits, such as transmons,  require nearly resonant resonators with dimensions scaling inversely with the frequency (as an illustration, a $\SI{1}{\mega\hertz}$ $\lambda/2$ coplanar cavity requires a $100$-m-long waveguide). 
Second, low-frequency systems are coupled to a hot thermal bath with which they exchange photons randomly, quickly turning pure quantum states into statistical mixtures. 

In recent years, significant progress has been made in overcoming these challenges. Notable contributions include the development of a $\SI{14}{\mega\hertz}$ heavy fluxonium qubit with a long coherence time and fast manipulation through fast-flux gates~\cite{Zhang2021}. Furthermore, operation of a fluxonium qubit dispersively coupled to a $\SI{690}{\mega\hertz}$ piezoelectric mechanical system was demonstrated earlier this year~\cite{Lee2023}. 

In this work, we demonstrate a fluxonium qubit with a transition frequency as low as \qubitfreq~MHz, achieving coherent operation and a charge sensitivity of $33~\mu$e/$\sqrt{\mathrm{Hz}}$, reflecting its potential for coupling with other devices operating in the MHz range. 
We achieve single-shot readout and direct preparation in the qubit state basis using sideband cooling, attaining a preparation fidelity above $\SI{97}{\percent}$. Based on this fidelity, we calculate an effective temperature of approximately $\SI{23}{\micro\kelvin}$. 
We also demonstrate direct resonant manipulation of the qubit state with a charge-drive as low as $5\cdot10^{-3}$ Cooper pairs. 
This value corresponds to a single-shot charge sensitivity of $10^{-2}$~e. 
In order to compare the sensitivity of our qubit-based detection scheme to other charge sensors, we accumulate statistical data through a cyclic qubit preparation and interrogation sequence. 
The charge sensitivity demonstrated here, at $\delta q \approx 33~\mu$e/$\sqrt{\mathrm{Hz}}$, rivals that of the most advanced transport-based devices~\cite{Korotkov1999, Angus2008, Lu2003, Schoelkopf1998, Cassidy2007, Volk2019, Gonzalez2015, Viennot2014, Brenning2006, Blencowe2000}, while maintaining intrinsic insensitivity to DC charge noise. 
The larger capacitance $C \sim 50$~fF of the superconducting island of our system results in an energy sensitivity, expressed in joules  per  hertz, $\delta q^2/ 2 C \sim 2.8\,\hbar.$  

The demonstrated charge sensitivity, combined with large gate capacitance demonstrated here 
are well-suited to explore quantum phenomena with low-frequency mechanical systems. 
For one, the frequency and electrode capacitance demonstrated in our work align with those found in cutting-edge vacuum-gap dispersive electromechanical systems~\cite{Seis2022}. 
Additionally, the single-shot charge sensitivity demonstrated here is sufficient to detect the zero-point motion of such a system placed in a DC-biased vacuum gap capacitor~\cite{Viennot2018}. 
Achieving the strong coupling between a low-frequency mechanical resonator and a superconducting qubit would enable to test the superposition principle in a regime where general relativity and quantum mechanics interplay~\cite{Gely2021}.

\section{Circuit design}

The heavy fluxonium circuit is shown on Fig.~\ref{fig:figure1}. The qubit itself is composed of a small Josephson junction (energy $E_J = \Phi_0^2/2 L_J$, $\Phi_0$ denoting the quantum of flux) shunted by a large capacitance to ground (capacitive energy $E_C = e^2/2 C$) and a superinductance (inductive energy $E_L = \Phi_0^2/2 L$) formed by 360 large Josephson junctions in series.
We ensure that each junction of the array has a negligible phase-slip rate by taking $E_{J, A}/E_p \gtrsim 3$, where $E_{J, A}/h = 65~$GHz is the Josephson energy of each array junction, and $E_p/h=17.9~$GHz is the junction plasma frequency~\cite{ManucharyanPhD}. In this regime the junction chain behaves as a linear inductor and the circuit Hamiltonian writes
\begin{equation}
    \label{eq:hamiltonian}
    \widehat H_Q = -E_J \cos\left( \hat \varphi - \varphi_\mathrm{ext} \right) + 4 E_C (\hat n - n_g(t))^2 + \frac{E_L}{2} \hat \varphi^2.
\end{equation}
In this equation, $\hat \varphi$ represents the superconducting phase across the junction, and $\hat n$ denotes its conjugate variable (the Cooper pair number),  $\varphi_\mathrm{ext} = 2\pi\Phi_{\mathrm{ext}}/\Phi_0$, where $\Phi_{\mathrm{ext}}$ stands for the magnetic flux threading the superconducting loop, and $n_g(t)$ is the offset charge on the capacitor pad. $\varphi_\mathrm{ext}$ can be controlled by an on-chip flux line, and $n_g(t)$ can be controlled by a capacitively coupled coplanar waveguide.  While the fluxonium spectrum is insensitive to a DC-charge offset~\cite{Manucharyan2009}, the main goal of this work is to evaluate the sensitivity of the qubit to a nearly resonant AC-charge modulation.

\section{Qubit spectrum}
The circuit operates in the heavy fluxonium regime, characterized by the two conditions $E_J \gg E_L$ and $E_J \gtrsim 10 E_C$. The first condition ensures that the potential experienced by the position-like variable $\hat \varphi$ consists of multiple wells with distinct minima. 
The second condition ensures that the lowest energy eigenstates are well localized within each well, with a small tunneling probability between neighboring wells. 
The magnitude of the tunneling rate $E_S/\hbar$ is exponentially suppressed as a function of $\sqrt{8 E_J/E_C}$, which relates the height of the potential barrier $2 E_J$ to the zero-point energy $\frac{1}{2} \sqrt{8 E_J E_C}$.

We denote $\ket{g}$ and $\ket{e}$ (resp.\ $\ket{f}$ and $\ket{h}$) as the fundamental (resp.\ first excited) states of the 2 lowest wells. 
Two families of transitions are observed in the two-tone spectroscopy of Fig.~\ref{fig:figure1}d: intra-well (or plasmonic) transitions, $\ket{g}\rightarrow\ket{f}$ and $\ket{e} \rightarrow \ket{h}$, that are only weakly dependent on the external flux $\varphi_\mathrm{ext}$, and inter-well transitions $\ket{e}\rightarrow\ket{f}$ and $\ket{g}\rightarrow\ket{h}$, that feature a linear dependence with $\varphi_\mathrm{ext}$.

Away from the symmetry points $\varphi_\mathrm{ext} \equiv 0 \,\,[\pi]$, the inter-well transition $\ket{g}\rightarrow\ket{e}$ is exponentially suppressed, acting as a selection rule that can be used to protect a microwave qubit against relaxation~\cite{Lin2017} (right inset of Fig.~\ref{fig:figure1}d).

At the flux-frustration point $\varphi_\mathrm{ext} = \pi$, the eigenstates undergo a transition, switching from localized modes around a single potential well to symmetric and anti-symmetric superpositions of these well-states. This transition results in a significant overlap of the flux wavefunctions, as evidenced by the magnitude of the flux matrix-element $|\langle g |\hat{\varphi} |e \rangle| \sim \pi$ (left inset of Fig.~\ref{fig:figure1}d). Importantly, at this point, the weakness of the tunneling element leads to a reduced qubit transition frequency $\omega_\mathrm{ge} = E_S/\hbar$. The value of $\omega_\mathrm{ge}$ can be tuned over several orders of magnitude by adjusting the circuit parameter $E_J/E_C$. In our specific case, we have chosen a transition frequency of \qubitfreq~MHz, which closely matches the oscillation frequency of existing macroscopic mechanical systems based on suspended membranes~\cite{Tsaturyan2017, Ivanov2020}. Notably, this frequency is approximately one order of magnitude lower than the lowest frequency ever reported using superconducting qubits~\cite{Zhang2021}.

\section{Sideband cooling}

With this low frequency, the qubit has almost equal ground and excited state populations at thermal equilibrium. Inspired by experiments with trapped ions~\cite{Diedrich1989} and optomechanical systems~\cite{Teufel2011}, we initialize the qubit in a pure state by driving the readout cavity with a detuned tone. 
By sweeping the reset tone frequency $\omega_p$ in the vicinity of the cavity resonance $\omega_\mathrm{R}/2\pi=5.64$~GHz, we observe two distinct processes at the sideband frequencies $\omega_\mathrm{R} \pm \omega_\mathrm{ge}$, corresponding to the transitions $\ket{g0}\rightarrow\ket{e1}$ and $\ket{e0}\rightarrow\ket{g1}$. 
More quantitatively, the qubit-resonator Hamiltonian can be linearized around the intracavity drive amplitude $\alpha$. For large drive amplitude and droping all terms rotating at the drive frequency, we arrive at the Hamiltonian (see Appendix~\ref{app:cooling})
\begin{equation}\label{eq:H4}
    \widehat H = \widehat H_\mathrm{Q} +  \hbar \Delta_\mathrm{R} \hat a^\dagger \hat a +  \hbar g \cos(\hat \varphi - \varphi_\mathrm{ext}) ( \alpha \hat a^\dagger + \alpha^* \hat a), 
\end{equation}
where $\hat a$ is the annihilation operator for photons in the readout cavity, $\Delta_\mathrm{R}=\omega_p - \omega_\mathrm{R}$ the pump-cavity detuning, and $\hbar g = E_J \varphi_\mathrm{zpf, R}^2$, with $\varphi_{\mathrm{zpf, R}}$  the zero-point fluctuations of the readout mode quantifying the energy-participation ratio of the cavity in the fluxonium junction. 
This Hamiltonian, expressed in a frame rotating at the drive frequency, describes the interaction between the fluxonium and an effective cavity mode of frequency $\Delta_\mathrm{R}$. 
When $\Delta_\mathrm{R}$ matches the frequency $\omega_\mathrm{eg}$ (respectively $-\omega_\mathrm{eg}$), the interaction reduces to the Jaynes-Cummings (respectively anti-Jaynes-Cummings) model between the cavity and the qubit. 
Furthermore, owing to the large cavity damping rate $\kappa/2\pi = 2.4$~MHz $\gg g |\alpha|$, the cavity field dynamics can be adiabatically eliminated, leading to the Purcell-like loss operators
\begin{equation}
    \label{eq:qubitDissipator}
    L^\pm = \frac{2 g |\alpha|}{\sqrt{\kappa}} \langle e | \cos(\hat \varphi - \varphi_\mathrm{ext}) | g\rangle \sigma^\pm,
\end{equation}
where the $\pm$ sign depends on the sideband addressed by the drive pulse $\Delta_\mathrm{R} \simeq \pm \omega_\mathrm{ge}$. 

At the flux-frustration point $\varphi_\mathrm{ext} = \pi$, the matrix element $\langle e | \cos(\hat \varphi - \varphi_\mathrm{ext}) | g\rangle$ cancels due to the opposite parity of the qubit wavefunctions.
Prior to the 10~$\mu$s reset pulse, we thus offset the flux by about $10^{-3} \cdot \Phi_0$ which corresponds to $\omega_{ge}/2\pi \simeq 10$ MHz, and we ramp it back to the frustration point afterwards. 
In order to avoid undesired mixing of qubit states caused by non-adiabatic effects~\cite{Zhang2021} while minimizing qubit decay, we have chosen a ramp duration of $2\,\mu$s.

The qubit population is then detected by standard cQED readout.
We were unable to directly resolve the qubit states $\ket{g}$ vs.~$\ket{e}$, due to a too small dispersive shift of the readout cavity. The qubit population was thus obtained by first mapping the population from $\ket{g}$ to $\ket{h}$, thanks to a $\pi$ Rabi pulse. The population in $\ket{h}$ is then measured by standard dispersive readout. The raw single-shot probability of detection are given by $P_g^{\mathrm{prep}~g}= 86.4 \%$ for a qubit prepared in $\ket{g}$ (resp. $P_e^{\mathrm{prep}~e}=92.8 \%$ for preparation in $\ket{e}$). By correcting for mislabelling and decay during readout (see Appendix~\ref{app:fidelity}), we infer a state preparation efficiency of 97.7~\% for qubit preparation in $\ket{g}$ and 97.7~\% for the preparation in $\ket{e}$.

\begin{figure}
\centering
    \includegraphics[width=0.47\textwidth]{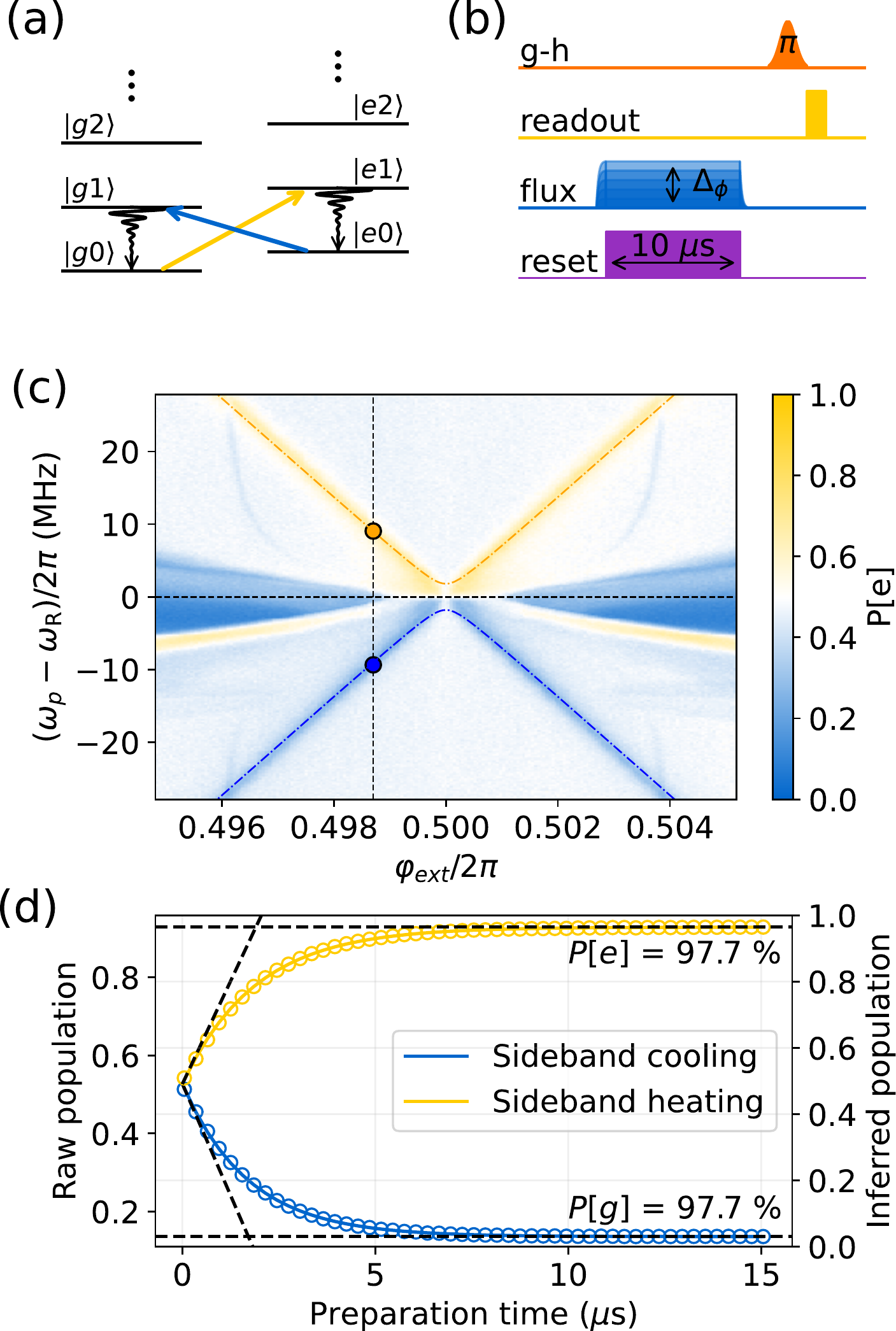}
    \caption{Sideband preparation of the fluxonium qubit. (a) Level diagram illustrating the sideband reset protocol:  The qubit is reset to either $\ket{g}$ or $\ket{e}$ by driving one of the sideband transitions: $\ket{e0} \rightarrow \ket{g1}$ or $\ket{g0} \rightarrow \ket{e1}$. The subsequent rapid decay of the cavity photon ensures a directional transition towards the desired qubit state. (b) Pulse sequence for the reset protocol. The flux bias (blue) is set to the target value within 2~$\mu$s. A reset tone (purple) is applied to the readout resonator port for 10~$\mu$s. The flux is then reset to $\varphi_\mathrm{ext} = \pi$ within 2~$\mu$s. The qubit state is read out by transferring the population from $\ket{g}$ to $\ket{h}$ (orange $\pi$-pulse) followed by single-shot dispersive readout in the eh manifold (yellow pulse). (c)
    Qubit population (color-scale) as a function of flux bias (x-axis) and reset pulse detuning (y-axis). The reset pulse power is adjusted to maintain a constant intracavity field at $\omega_\mathrm{R} \pm \omega_\mathrm{ge}(\varphi_\mathrm{ext})$. The horizontal dashed line indicates the readout frequency. The nominal working points for $\ket{e}$ and $\ket{g}$ preparation are denoted by the orange and blue dots, respectively. The orange and blue dotted-dashed lines represent the predicted frequencies based on the Hamiltonian parameters in Fig.~\ref{fig:figure1}.
    (d) Final qubit population as a function of reset pulse duration (other parameters are the nominal parameters indicated in panel (c)). The population extracted from single-shot readout distributions is shown on the left axis, while the right axis displays the population corrected for decay and mislabeling during readout (see Appendix~\ref{app:fidelity}). Exponential fits to the data (yellow and blue lines) yield preparation times of 1.9 $\mu$s ($\ket{e}$) and 1.7 $\mu$s ($\ket{g}$) with final occupations of 97.7 \% and 97.7 \%, respectively.
    \label{fig:figure2}}
\end{figure}

\begin{figure*}[t]
    \centering
    \includegraphics[width=0.99\textwidth]{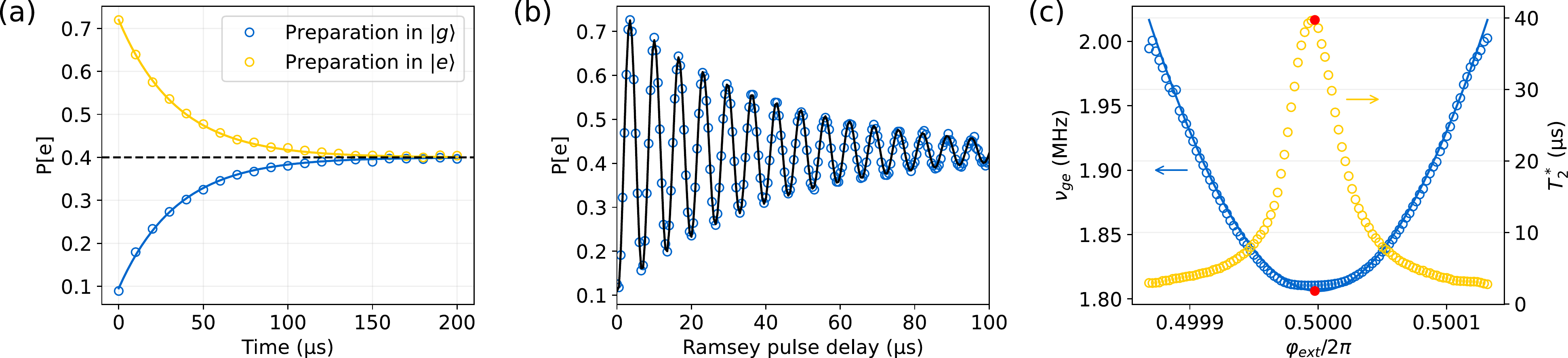}
    \caption{Qubit coherence. (a) Energy relaxation time ($T_1$) measured at the flux point $\varphi_\mathrm{ext} = \pi$: the raw qubit population is plotted as a function of delay time after preparation in $\ket{e}$ (yellow dots) or $\ket{g}$ (blue dots) as the qubit relaxes towards a thermal mixture. A common exponential fit yields $T_1 = 34\,\mu$s. The apparent imbalance in the equilibrium population (dashed line) is due to the residual decay of the intermediate state $\ket{f}$ used for readout. (b) Ramsey experiment after state preparation in $\ket{g}$. The exponential decay of the Ramsey fringes (black line) yields $T_2^* = 39.7\,\mu$s, and a transition frequency of \qubitfreq~MHz.  (c) $T_2^*$ (yellow curve, right axis) and transition frequency (blue curve, left axis) as a function of flux bias. The points highlighted in red correspond to the data presented in panel (b). The blue solid line is the predicted qubit frequency, for the Hamiltonian parameters given in Fig.~\ref{fig:figure1}.
    \label{fig:coherence}}
\end{figure*}

\section{Qubit coherence}
After having established this preparation process, we first investigate the energy relaxation of the $\ket{e}$ and $\ket{g}$ fluxonium states, towards a thermal state: Fig.~\ref{fig:coherence}a displays the qubit population versus the delay time after preparation in either $\ket{e}$ or $\ket{g}$. 
When a qubit interacts with a thermal environment of occupation $n_\mathrm{th}$, it experiences two loss channels, described by the operators  $\sqrt{\Gamma_\uparrow} \hat \sigma^+$ and $\sqrt{\Gamma_\downarrow} \hat \sigma^-$, where $\Gamma_\uparrow \propto n_\mathrm{th}$ and  $\Gamma_\downarrow \propto n_\mathrm{th} + 1$. 
In the case of low-frequency transitions, such as the $\{\ket{e}, \ket{g}\}$ manifold, the large environmental occupation $n_\mathrm{th} \sim k_B T_\mathrm{eg}/\hbar \omega_\mathrm{ge}$, with $T_\mathrm{eg}$ being the environmental temperature associated to the \qubitfreq~MHz transition, results in $\Gamma_\uparrow \approx \Gamma_\downarrow \equiv \Gamma$, leading to an exponential relaxation  towards the statistical mixture $\rho^\mathrm{th} = (\ket{e} \bra{e} + \ket{g} \bra{g})/2$ at a rate $2 \Gamma$. 
By fitting exponential curves to the data of Fig.~\ref{fig:coherence}a, we obtain $T_1 = 1/2 \Gamma = 34~\mu$s. 

As the qubit frequency explored in the current work extends well below the values reported in the literature so-far~\cite{Zhang2021}, it is important to determine whether the qubit transition couples with a thermal environment or is primarily constrained by technical noises (e.g. 1/f charge noise). 
To examine this, we conducted $T_1$ measurements similar to those shown in Fig.~\ref{fig:coherence}a, while varying the cryostat base temperature (see Appendix \ref{app:t1_vs_t}). 
In order to compare this relaxometry measurement conducted on the \qubitfreq~MHz qubit transition with the environmental temperature experienced by GHz-frequency transitions, we used the residual population in the $\{\ket{f}, \ket{h}\}$ manifold as a local probe for the qubit temperature $T_\mathrm{ef}$.  
Although the built-in temperature sensor of the cryostat indicates a minimal temperature of 7 mK, we have found that the circuit only thermalizes to $T_\mathrm{ef}\approx 59$~mK. Nevertheless,  we observe a nearly linear relationship between $\Gamma$ and $T_\mathrm{ef}$ down to $T_\mathrm{ef} \approx 100~$mK, suggesting comparable noise temperatures $T_\mathrm{ef}$ and $T_\mathrm{eg}$ for the 1.8 MHz and 3.7 GHz transitions. 
This observation indicates that, despite its ultra-low operational frequency, our qubit is marginally impacted by 1/f noise. 
This outcome stands in contrast with recent studies on frequency-tunable fluxonium~\cite{Sun2023} and may be attributed to the small superconducting loop area used in our circuit, limiting the influence of flux-noise.

Finally, we probe the qubit dephasing time, denoted as $T_2^*$, as a function of external flux. To achieve this, we conducted Ramsey sequences on the $\ket{g}\rightarrow \ket{e}$ transition. As seen in Fig.~\ref{fig:coherence}c, the coherence time reaches its maximal value of approximately 40~$\mu$s at the flux frustration point, $\varphi_\mathrm{ext} = \pi$. 
Indeed, as shown on Fig.~\ref{fig:coherence}c, the qubit frequency is to first order insensitive to fluctuations in the external magnetic flux at this point. 
The Ramsey fringe measurement at $\varphi_\mathrm{ext} = \pi$ is depicted in Fig.~\ref{fig:coherence}b. 
Notably, the measured coherence is not too far from the upper limit of $2 T_1$, suggesting a pure dephasing rate of $\Gamma_\phi = 1/2 T_1 - 1/T_2^* = (97~\mu$s$)^{-1}$.

\section{AC-Charge sensitivity of the fluxonium qubit}

In the following, we evaluate the sensitivity of the fluxonium to a nearly resonant AC-charge drive. 
We delve first into the theoretical advantages of the fluxonium qubit over other qubit implementations, before introducing a practical scheme for the experimental detection of weak charge modulation.

\subsection{Advantage of the heavy-fluxonium over other capacitively-shunted qubits}

In this section, we aim to maximize the Rabi rate for a single-mode qubit subjected to a nearly-resonant offset charge of fixed oscillation amplitude $N_\mathrm{drive}$ and frequency $\omega_d$. This thought experiment will provide a clearer understanding of why the heavy-fluxonium holds an advantage over other capacitively-shunted qubits.

Consider a single-mode qubit with a capacitive energy given by $4 E_C (\hat n - n_g(t))^2$, which interacts with a classical offset charge $n_g(t) = N_\mathrm{drive} \cos(\omega_d t)$. For small charge modulations $N_\mathrm{drive} \ll 1$, the Hamiltonian can be linearized. In a frame rotating at the drive frequency, it writes
\begin{equation}
    \hat H_\mathrm{int}
    = - 8 N_\mathrm{drive} E_C  \langle e | \hat n | g \rangle  \hat \sigma_x.
\end{equation}
Using the relation between charge and flux matrix elements, namely $8E_C|\!\!\braket{e | \hat n | g}\!\!| = \omega_\textrm{eg}|\!\!\braket{e | \hat \varphi | g}\!\!|,$ we derive the Rabi frequency
\begin{equation}
\label{eq:rabi}
\Omega_\textrm{r} = 2 N_\mathrm{drive}\omega_\textrm{ge}|\!\braket{e | \hat \varphi | g}\!|.
\end{equation}
In a resonant coupling scenario, where the drive frequency $\omega_d$ is imposed by the resonance of an auxiliary system to probe, the qubit frequency needs to fulfill $\omega_\mathrm{ge}=\omega_d$. In such a situation,  maximizing the third factor $|\!\braket{e | \hat \varphi | g}\!|$ is crucial.  
Indeed, only this term depends on the specifics of the qubit implementation, while the first two terms $N_\mathrm{drive}$ and $\omega_\textrm{ge}$ are characteristics of the auxiliary system to be detected. 
For instance, in cQAD, the frequency $\omega_d$ is set by the mechanical resonance frequency, whereas the amplitude $N_\mathrm{drive}$ depends on the details of the mechanical-electrical transduction. 
Consider the scenario of a silicon nitride membrane, which is a promising candidate for testing Penrose gravitational collapse due to its long coherence time and large zero-point fluctuations~\cite{Gely2021}. In this case, we expect an AC-charge modulation of $N_\mathrm{drive} \sim 10^{-2}$ at a resonance frequency of $\omega_d/2 \pi = \Omega_m/2 \pi \approx 2$~MHz (see Appendix~\ref{app:membrane}).

While the matrix element $|\!\langle e | \hat \varphi | g \rangle\!|$ is typically suppressed exponentially in the heavy fluxonium regime,  a radically different scenario emerges at the flux-frustration point. Here, the wavefunctions recover a large overlap $|\!\langle e | \hat \varphi | g \rangle\!| \sim \pi$. This value compares favorably with weakly anharmonic qubits, where $|\!\langle e | \hat \varphi | g \rangle\!| \sim (2 E_C/E_J)^{1/4} \ll 1$, or even the Cooper-pair box $|\!\langle e | \hat \varphi | g \rangle\!|\sim 4 E_C/E_J\sim 1$. 
In essence, the unique characteristics of fluxonium eigenstates at the flux-frustration point ---  manifesting as Schr\"odinger cat-like superpositions of persistent current states --- endow it with a larger charge sensitivity compared to a transmon or Cooper-pair box operating at the same transition frequency.
\begin{figure}
    \centering
    \includegraphics[width=0.48\textwidth]{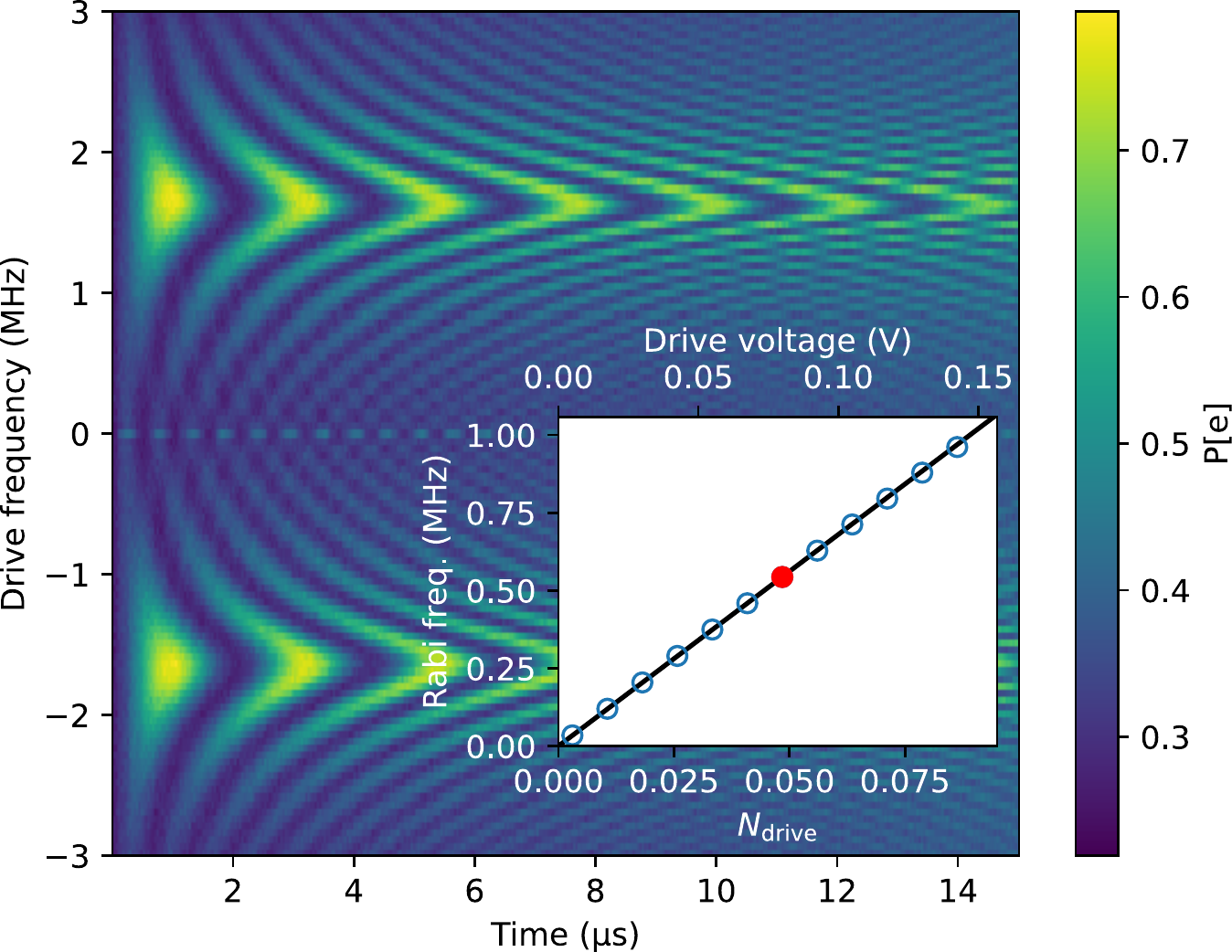}
    \caption{Direct Rabi manipulation of the radio frequency qubit transition. after initial preparation in $\ket{g}$, the flux is reset to $\varphi_\mathrm{ext} = \pi$, and the qubit is driven via the charge port with a MHz pulse of variable duration and frequency. The final qubit population is read out using the technique described in Fig.~\ref{fig:figure2}. The negative frequency part of the graph is here to highlight the validity range of the rotating-wave approximation. The inset shows the Rabi frequency for a resonant drive at \qubitfreq~MHz, extracted from a sinusoidal fit, as a function of the drive voltage amplitude (upper horizontal axis). A linear fit of Eq.~\eqref{eq:rabi} to the data provides the lower horizontal axis calibration, where the drive amplitude is expressed in Cooper-pairs on the fluxonium electrode. The red dot is obtained for the parameters of the main figure.  
    \label{fig:chevron}}
\end{figure}

\begin{figure*}[t]
    \centering
    \includegraphics[width=\textwidth]{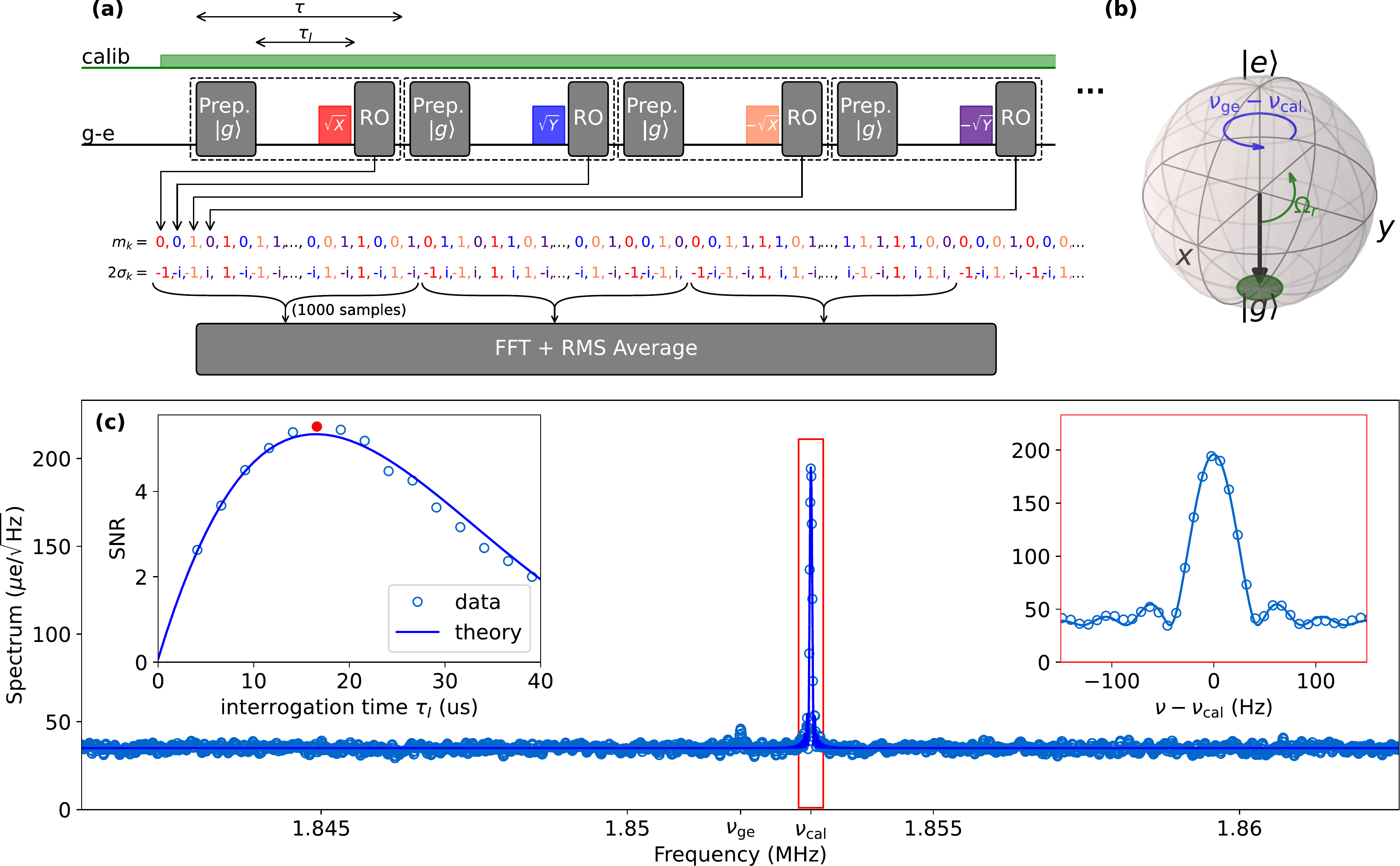}
    \caption{AC-charge sensing. (a) A weak monochromatic charge-drive (also referred to as calibration tone) is detected thanks to a repeated pulse sequence: the qubit is prepared in $\ket{g}$ (black arrow in the Bloch sphere (b)). After interacting for a time $\tau_I$ with the tone, a partial information on the qubit state is obtained by performing a $\pi/2$ pulse in one of the 4 directions ${+X,  +Y, -X, -Y}$, followed by a qubit state readout in the eg basis. From the measurement samples $m_k \in$ \{0, 1\}, a complex telegraphic signal $\sigma_{k} = i^k (m_k - 1/2)$ is constructed. The noise spectrum centered around the qubit frequency is estimated by the Bartlett's-method, with periodograms of 1000 non-overlapping consecutive samples. (c) The estimated noise spectrum presents a residual-bandwidth-limited peak at the calibration tone frequency $\nu_\mathrm{cal} = 1.853$~MHz. Red inset: zoom on the calibration peak and sinus-cardinal fit (solid line). Left and right insets: signal-to-noise-ratio (SNR) for the calibration peak as a function of interrogation time $\tau_I$ and calibration peak amplitude respectively. The red dots in the insets correspond to the parameters used in the main graph of panel (c). The solid lines are the results of an analytic model taking into account the evolution of the qubit during the interrogation time. Signal cancellation occurs when the calibration tone amplitude is a multiple of that of a pi-pulse. The spectrum in (c) is calibrated using the known variance of the calibration tone.}
    \label{fig:spectrum}
\end{figure*}

\subsection{Rabi oscillations of the qubit transition}

In Fig.~\ref{fig:chevron}, we directly drive the qubit, biased at $\varphi_\mathrm{ext} = \pi$, with a MHz pulse on the charge drive. We observe a Rabi oscillation pattern with maximum contrast for $\omega_d = \omega_\mathrm{ge}$. 
The inset shows the Rabi frequency $\Omega_r/2\pi$ for a resonant drive at \qubitfreq~MHz. 
As expected from relation \eqref{eq:rabi}, we observe a linear dependence of the Rabi oscillations with the drive amplitude, up to $\Omega_r/2\pi \sim$ 1 MHz. For larger amplitude of the drive, the rotating wave approximation breaks down as $\Omega_r$ approaches $\omega_\mathrm{ge}$, leading to a deformed pattern with reduced contrast at the resonance drive condition. 
We use relation \eqref{eq:rabi} to relate the voltage amplitude on the digital-to-analog converter to the equivalent number of Cooper-pairs $N_\mathrm{drive}$ on the fluxonium electrode. We also deduce from this relation the minimum charge modulation required to observe coherent Rabi oscillations 
\begin{equation}
    N_\mathrm{min} = \frac{2 \pi}{|\!\!\braket{e | \hat \varphi |g}\!\!| \omega_\mathrm{ge} T_1} \approx 5 \cdot 10^{-3}.
\end{equation}
The ability to manipulate the qubit state with less than one percent of a Cooper-pair shows the extreme sensitivity of the fluxonium to a resonant AC-charge modulation. 
For instance, this value would be sufficient to reach the strong-coupling regime with a DC-biased mechanical membrane in a resonant coupling scenario (see Appendix~\ref{app:membrane}). 

The aforementioned value of $5\cdot 10^{-3}$ Cooper pairs corresponds to a single shot charge sensitivity of $10^{-2}$~e.  However, through the implementation of quantum sensing protocols, like those routinely used in nitrogen-vacancy-center magnetometry~\cite{Bonato2016} and similar methodologies~\cite{Polino2020}, we are able to accrue substantial statistical data. This allows us to measure charge sensitivity within a one second integration period and subsequently compare these findings with other charge sensing methods.

\subsection{Frequency-resolved AC-charge sensitivity} 
In a quantum sensing experiment, we can leverage the ability to swiftly prepare and read out the qubit state to detect a weak charge signal through repeated interaction with the two-level system. This involves preparing the qubit in $\ket{g}$, after which it interacts  for an interrogation time $\tau_I$ with the weak continuous signal to be detected (referred to as the ``calibration tone'' henceforth), of frequency $\omega_\mathrm{cal}$, applied to the charge port.
For weak enough calibration tone, the Bloch-vector undergoes a small rotation away from the south pole. 
We then probe this displacement by mapping the transverse component of the Bloch-vector to the $\sigma_z$ basis with a $\pi/2$ pulse, before performing a single-shot readout of the qubit in the $\ket{g}$, $\ket{e}$ basis. 
In this scheme, the probability to detect the qubit in $\ket{e}$ slightly deviates from $1/2$, by an amount that depends on the phase and amplitude of the calibration tone. 
Furthermore, the mismatch $\Delta$ between the calibration tone and qubit frequencies gives rise to a shot-to-shot rotation of the Bloch-vector by an angle $\theta_k = k \Delta \tau$, where $k$ is the repetition index and $\tau$ the repetition period of the experiment. 
Even though each measurement result $m_k \in \{0, 1\}$ only contains one bit of information, the complete measurement record $\{m_k\}_{0 \le k <N_\mathrm{tot}}$  can be used to reconstruct the 
spectrum of the charge modulation by the periodogram method~\cite{Alan1999}. 

Performing the $\pi/2$ rotation along a unique axis would lead to an ambiguity between positive and negative detuning $\Delta$. We thus perform the qubit rotations along an axis picked up sequentially in the set ($+X$, $+Y$, $-X$, $-Y$). 
This ensures a non-ambiguous correspondence between discrete and continuous time frequencies over the interval $[-\Omega_\mathrm{Ny}/2, +\Omega_\mathrm{Ny}/2]$, where $\Omega_\mathrm{Ny}$ is the Nyquist angular frequency $\Omega_\mathrm{Ny} = \pi / \tau$ (see Appendix~\ref{app:spectrum}). 
The charge-noise spectrum over this interval is then reconstructed by performing fast-Fourier-transforms over adjacent windows of $N=1000$ consecutive samples.  Fig.~\ref{fig:spectrum}c shows an example of such an experimentally reconstructed spectrum. 
The calibration tone is visible as a sinus-cardinal-shaped peak, centered around $\omega_\mathrm{cal}$ and of width $\Omega_\mathrm{RBW} = 2\pi/N \tau$. This value is the residual bandwidth of our quantum spectrum analyzer, and it can be tuned by adjusting the window length $N$. 
The spectrum is normalized in units of elementary charge e$/\sqrt{\mathrm{Hz}}$ using the known amplitude of the calibration tone, as determined from the linear fit of Fig.~\ref{fig:chevron}. 

The calibration peak sits on a flat noise background, which is attributable to the sampling noise of the quantum sensor~\cite{Degen2017}. 
An analytic model for the signal-to-noise ratio (SNR) as a function of the experimental parameters has been derived (see Appendix~\ref{app:spectrum}) and shows good agreement with the measured data (see Fig.~\ref{fig:spectrum}c). 
Qualitatively, the SNR increases linearly for $\tau_I \ll T_1$, as the initial Bloch-vector accumulates a transverse component $2 |\!\braket{\sigma}\!| = \Omega_r \tau_I$.  
On the other hand, due to the interaction with the thermal bath, the Bloch-vector relaxes eventually towards the origin of the Bloch sphere such that the SNR vanishes for $\tau_I \gg T_1$. 
In practice, around the optimal value $\tau_I \sim 20$~$\mu$s, the detector achieves a noise-level as low as $33 ~\mu$e$/\sqrt{\mathrm{Hz}}$. 
This value approaches that of the most sensitive electrometers such as the radiofrequency quantum point contact (rf-QPC)~\cite{Reilly2007, Cassidy2007} or the radiofrequency single-electron transistor~\cite{Schoelkopf1998, Lu2003}. 
Yet, these transport-based sensors are very different in nature from the current qubit-based quantum protocol. 
The shunt-capacitor on which the charge is detected in our system is typically 2 orders of magnitude larger than the superconducting islands employed in those systems~\cite{Schoelkopf1998, Cassidy2007}. 
This is of utmost practical importance when it comes to connecting the sensor to an auxiliary quantum system. 
As an example, when trying to detect the charge-modulation of an electromechanical system such as~\cite{Seis2022}, the 50~fF capacitor of the vacuum-gap system would perfectly match the value employed in this work, whereas traditional sensors would suffer a large dilution of the signal. 
The challenge of detecting extremely small charge signals while maintaining a large island capacitance is more directly captured by the energy sensitivity~\cite{Schoelkopf1998}
$\delta q^2/2 C \approx 2.8~\hbar$ which is below the sensitivity of any other charge detectors operating at MHz frequencies. 
Furthermore, in stark contrast with transport-based measurements, featuring a flat frequency response from DC to several tens of MHz, our resonant detector features a narrow frequency response around the qubit frequency, the full bandwidth being given by $\Omega_\mathrm{full} = 2 \pi/\tau_I \sim 50$~kHz (see Appendix~\ref{app:spectrum}). 
This peculiar frequency response is highly advantageous when coupling the fluxonium to a nearly resonant system, as it guarantees perfect immunity to low-frequency environmental charge noise while maximizing charge sensitivity at the MHz region of interest.

\section{Conclusion}

In conclusion, we have demonstrated high-fidelity preparation, manipulation and single-shot readout of a heavy-fluxonium qubit with a transition frequency as low as \qubitfreq~MHz. 
To the best of our knowledge, this is the lowest frequency reported so far for a superconducting qubit.
As demonstrated in earlier work~\cite{Zhang2021}, this circuit represents a realistic alternative to the transmon in a quantum computing architecture. 
Our work furthermore demonstrates the potential of this circuit in sensing experiments. This can be routed from the peculiar frequency response of the circuit which filters efficiently the environmental noise at audio frequency while being maximally sensitive at the resonant qubit frequency in the MHz range. 
The high charge sensitivity combined with the large capacitive shunt demonstrated in this work opens up avenues in hybrid circuits, where the fluxonium can be used as a resonant probe to manipulate other physical systems. 
As an example, we show (Appendix~\ref{app:membrane}) that  the coherence time and electric dipole achieved in the current work are sufficient to attain the strong-coupling regime in an hybrid electromechanical system involving a DC-biased nanomechanical resonator.

\section{Acknowledgements}

The authors acknowledge support from ANR project MecaFlux (ANR-21-CE47-0011), the CryoParis project from the Région Ile-de-France DIM Sirteq, and the HyQuTech project from the Emergence Sorbonne Université program. This project
has received funding from the European Research Council (ERC) under the European Union’s Horizon 2020 research and innovation programme (grant agreements No.
851740). KG is funded by the Quantum Information Center Sorbonne (QICS) PhD program, HP is funded via the CNRS/UArizona joint PhD program. EF acknowledges funding from the European Research Council under grant no. 101042315 (INGENIOUS). AS acknowledges support from ANR project Hamroqs and the Plan France 2030 through the PEPR NISQ2LSQ project (ANR-22-PETQ-0006). LN acknowledges support from ANR QFilters (ANR-18-JSTQ-0002).

\appendix

\section{Micro-fabrication}
The large circuit parts, \emph{i.e.}, the coplanar wave-guide resonator, the flux line, the charge-drive electrode, and the fluxonium coplanar-capacitor, were fabricated with standard UV laser lithography: starting with a 280-$\mu$m-thick silicon (100) wafer with a resistivity of 20~$k\Omega.$cm, and coated with a 150~nm-thick layer of niobium (Nb), we spin-coat   S1805 positive resist, and bake it at 115~\textdegree C for 1~min.  The resist is then  exposed to UV light with a dose of 100 mJ/cm$^2,$ and developed using MF-319. Nb is etched using reactive ion etching (RIE) with a SF$_6$ plasma. Any remaining resist is finally removed with acetone in an ultrasound bath at 50~\textdegree C for 15 min, rinsed in IPA and dried.\\

The small Josephson junction and the numerous large junctions of the superinductor were fabricated using the Dolan-bridge technique. First, the sample is spin-coated with MMA EL13 at 4000~RPM, and subsequently baked for 1~min at 195~\textdegree C. The sample is then spin-coated with PMMA A3 at 5000~RPM and baked for 30~min at 195~\textdegree C. Electron-beam lithography with a dose of 280~$\mu$C/cm$^2$ is used to create the free-standing bridges on the MMA-PMMA bilayer. The development is performed with a mixture of IPA and DI water at a temperature of 6~\textdegree C for 90~s. The Al/AlO$_x$/Al junctions are fabricated by evaporation of a 31~nm-thick aluminum (Al) layer with a $-$22\textdegree~angle, followed by oxidization with a mixture~9:1 of argon and O$_2$, at $200$~mbar and for 12~min. Finally, a $100$~nm-thick Al layer is evaporated at $+$22\textdegree~angle. A lift-off is then performed in a NMP bath at 80~\textdegree C for 20~min. The sample is then dried after rinsing with acetone and IPA.  

\section{Experiment schematic}

\begin{figure}[t]  \includegraphics[width=0.48\textwidth]{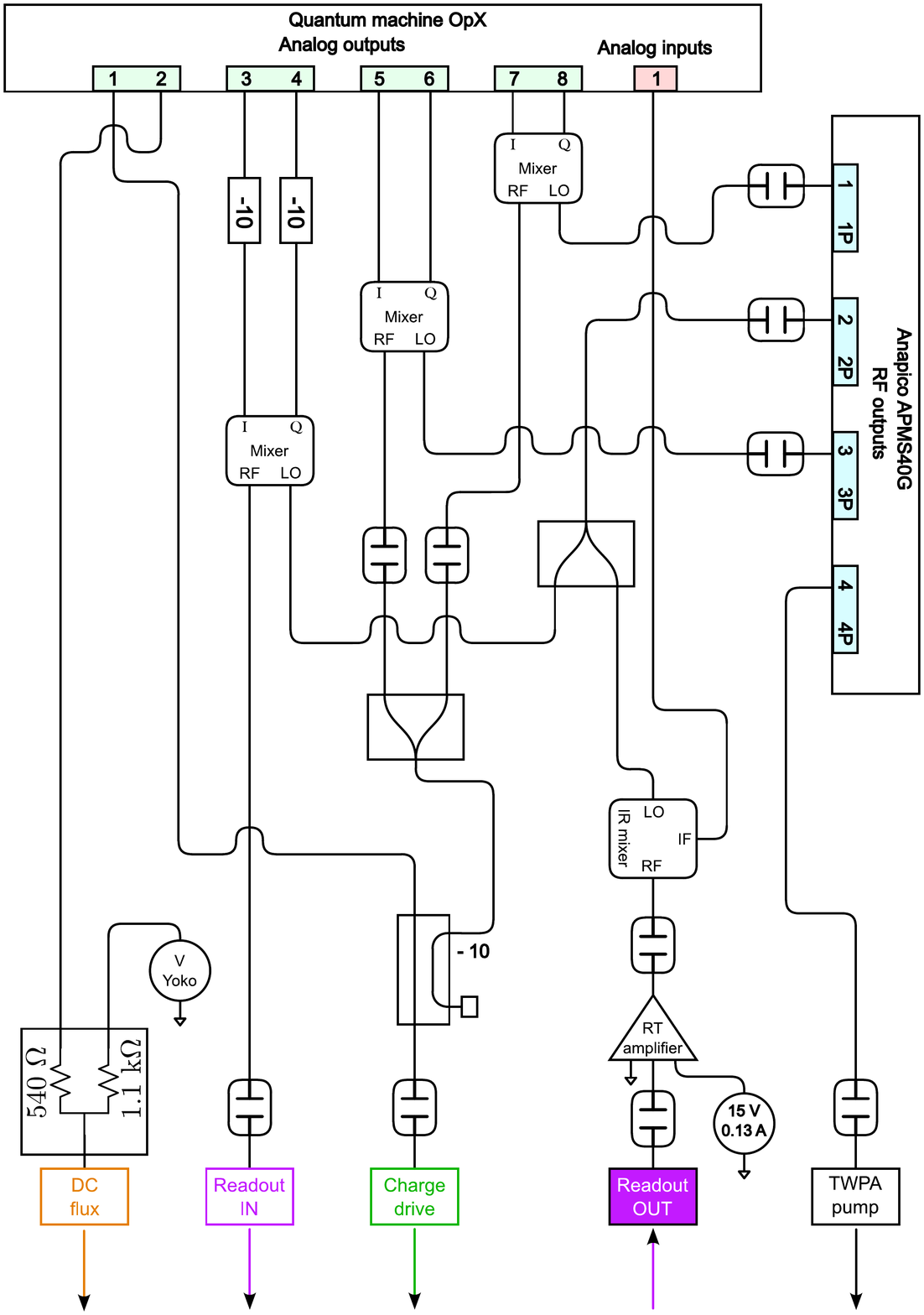}
    \caption{Room temperature RF- and DC-circuitry. }
    \label{fig:rtwiring}
\end{figure}

\begin{figure}[t]  \includegraphics[width=0.48\textwidth]{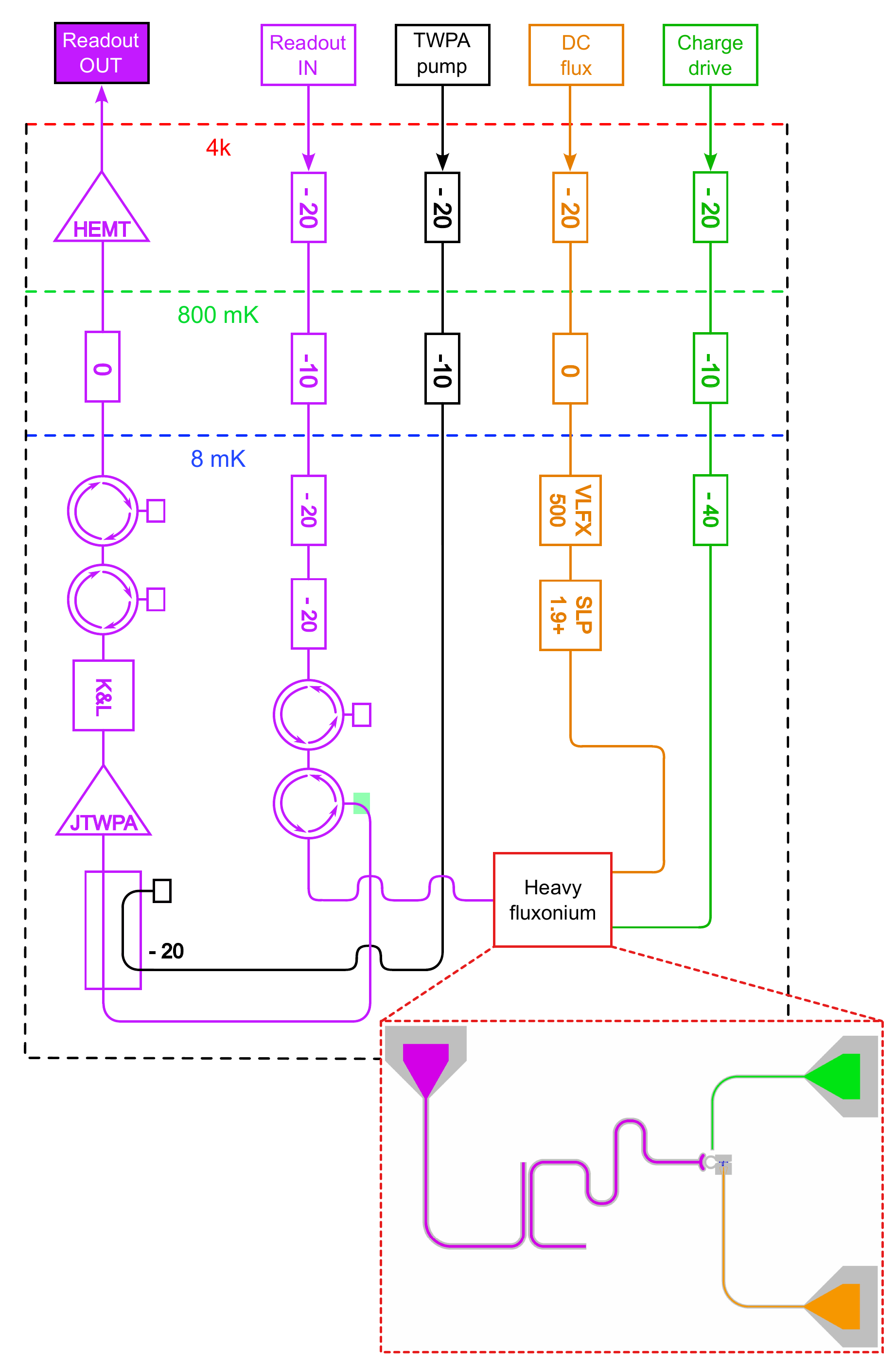}
    \caption{Cryogenic RF- and DC-circuitry. The color code matches that of Fig.~\ref{fig:figure1}.}
    \label{fig:cryowiring}
\end{figure}

The room-temperature and cryogenic RF- and DC-connections are depicted on Fig.~\ref{fig:rtwiring} and \ref{fig:cryowiring} respectively. 
A fast data acquisition system (OPX, Quantum Machine) is used in combination with a microwave source (Anapico APMS40G) to generate the RF- and microwave pulses. 
The magnetic flux control is obtained by combining a stabilized voltage source (Yokogawa 7651) with a fast analog output of the OPX. 
this setup allows us to scan the flux-bias over more than $\Phi_0$, while enabling fast control over a range of approximately $3\cdot10^{-2} \cdot \Phi_0$ with a sub-$\mu$s time resolution. 

\section{Sideband cooling}
\label{app:cooling}
The starting point of our analysis is the Hamiltonian of the fluxonium, including the driven readout cavity. The Hamiltonian is expressed in the normal mode basis, as obtained by diagonalizing the classical equations of motion for $E_J=0$~\cite{Smith2016prb}.  The normal modes are labeled ``R'' for \emph{readout-like} and ``Q'' for \emph{qubit-like}. The Hamiltonian writes
\begin{align}
    \widehat H = & \hbar\omega_\mathrm{Q} \hat c^\dagger \hat c + \hbar\omega_\mathrm{R} \hat a^\dagger \hat a + \hbar(\hat a \varepsilon_d^* e^{i\omega_p t} + \hat a^\dagger\varepsilon_d e^{- i\omega_p t}) \\
    & -E_J \cos(\hat \varphi_\mathrm{Q} +\hat \varphi_\mathrm{R}  - \varphi_\mathrm{ext}),
\end{align} where $\hat \varphi_\mathrm{R} = \varphi_{zpf,R} (\hat a + \hat a^\dagger),$ and $\hat \varphi_\mathrm{Q} = \varphi_{zpf,Q} (\hat c + \hat c^\dagger),$ represent the normal mode position-like operators in the absence of the Josephson term ($E_J=0$). The bosonic annihilation operators for the qubit and cavity modes are denoted by $\hat c$ and $\hat a$, with respective frequencies $\omega_\mathrm{Q}$ and $\omega_\mathrm{R}$. The zero-point fluctuations of the readout and qubit modes, as seen by the Josephson junction, are given by $\varphi_\mathrm{zpf,R}$ and $\varphi_\mathrm{zpf,Q}$. Note that the participation ratio of the resonator mode in the Josephson junction is very small~\cite{Smith2016prb}, such that $\varphi_\mathrm{zpf,R}\ll 1$.

To account for photon loss in the readout resonator, we model the dynamics with a master equation
\begin{equation}\label{eq:FULLsystem}
\tfrac{d}{dt} \hat \rho = -i [\widehat{H},\, \hat \rho] + \kappa \mathcal{D}_{\hat a}(\hat \rho) \; , 
\end{equation} where the Lindbladian writes $\mathcal{D}_{\hat a}( \hat \rho) = \hat a \hat \rho \hat a^\dagger - \frac{1}{2}[\hat a^\dagger \hat a \hat \rho + \hat \rho \hat a^\dagger \hat a]$.

In the following, we demonstrate how Eq.~\eqref{eq:FULLsystem} simplifies to the effective qubit dissipative dynamics, as represented by the loss operator in Eq.~\eqref{eq:qubitDissipator}. 
We proceed by going in a frame rotating at the drive frequency $\omega_p$, and displaced around the mean amplitude of the cavity field $\alpha$, by doing the substitution $\hat a \rightarrow (\alpha + \hat a) e^{- i\omega_p t}.$ The steady state value $\alpha$ is chosen such that it cancels the three drift terms
\begin{align}
\widehat H_1 = & (\hbar\kappa/2i)\left(\alpha^*\hat a - \alpha \hat a^\dagger\right)\\
\widehat H_2 = & \hbar \Delta \left(\alpha\hat a^\dagger + \alpha^*\hat a\right)\\
\widehat H_3 = & \hbar\left(\epsilon_d^*\hat a + \epsilon_d\hat a^\dagger\right),
\end{align} where $\widehat H_1$ is an effective Hamiltonian dynamics stemming from the expression of the Lindbladian in the displaced frame, $\widehat H_2$ comes from the linearization of the term $\hbar\Delta_\mathrm{R}\hat a^\dagger\hat a$ in the rotating frame Hamiltonian, with $\Delta_\mathrm{R} = \omega_p -\omega_\mathrm{R}$ being the drive detuning, and $\widehat H_3$ is the drive term. 
We thus obtain the value of $\alpha$:
\begin{equation}
\alpha = \frac{-\epsilon_d}{\Delta_\mathrm{R}+i\kappa/2}.
\end{equation}
The Hamiltonian in the displaced frame becomes 
\begin{align}
\widehat H = & \hbar\omega_\mathrm{Q} \hat c^\dagger \hat c + \hbar\Delta_\mathrm{R} \hat a^\dagger \hat a\\
& -E_J\cos\left(\hat \varphi_{\mathrm{Q}} +\tilde{\varphi}_{\mathrm{R}} -\varphi_{\mathrm{ext}}\right).\nonumber
\end{align}
where $\tilde{\varphi}_{\mathrm{ R}} = \varphi_{\mathrm{zpf, R}}\left(\alpha +\hat a\right)e^{- i\omega_p t} +\varphi_{\mathrm{zpf, R}}(\alpha^* +\hat a^\dagger)e^{i\omega_p t}$ is the resonator coordinate in the new frame.
Since $\varphi_{\mathrm{zpf, R}}|\alpha| \ll 1$, we Taylor expand this expression to second order with respect to $\tilde{\varphi}_{\mathrm{R}}$
\begin{align}
\label{eq:HTaylor}
\widehat H = & \hbar\omega_\mathrm{Q} \hat c^\dagger \hat c + \hbar\Delta_\mathrm{R} \hat a^\dagger \hat a -E_J\cos\left(\hat \varphi_{\mathrm{Q}} -\varphi_{\mathrm{ext}}\right) \\
& + E_J \sin\left( \hat \varphi_{\mathrm{Q}}-\varphi_{\mathrm{ext}}\right) \cdot \tilde{\varphi}_{\mathrm{ R}} \nonumber\\
&+\frac{E_J}{2} \cos\left( \hat \varphi_{\mathrm{Q}}-\varphi_{\mathrm{ext}}\right) \cdot \tilde{\varphi}_{\mathrm{ R}}^2. \nonumber
\end{align}
The first line of Eq.~\eqref{eq:HTaylor} corresponds to the resonator and unperturbed qubit Hamiltonian, the second line, which corresponds to the first order Taylor expansion, can be safely neglected as it only consists of terms rotating at $\pm\omega_p$. On the other hand, the third line, corresponding to the second order Taylor expansion, reduces to 
$\cos\left( \hat \varphi_{\mathrm{Q}}-\varphi_{\mathrm{ext}}\right) \cdot \varphi_\mathrm{zpf, R}^2( \alpha \hat a^\dagger + \alpha^*\hat a)$, once fast rotating terms have been neglected, and linearizing for $\alpha \gg \hat a, \hat a^\dagger$. We thus obtain Eq.~\eqref{eq:H4}, with 
\begin{align}
    \widehat H_\mathrm{Q} &= \hbar \omega_\mathrm{Q} \hat c^\dagger \hat c\nonumber\\
    &  - E_J \, \left(1-(\tfrac{2 \pi \Phi_{\mathrm{zpf,R}}}{\Phi_0})^2(|\alpha|^2+\tfrac{1}{2})\right)\, \cos(\hat \varphi_Q- \varphi_\mathrm{ext}),\nonumber\\
    &\simeq \hbar \omega_\mathrm{Q} \hat c^\dagger \hat c - E_J \cos(\hat \varphi - \varphi_\mathrm{ext}).
\end{align}
In the last expression, we have used $\varphi_{\mathrm{zpf, R}}|\alpha| \ll 1$, and $\hat \varphi_Q \approx \hat \varphi$ since the participation ratio of the readout resonator in the junction is small. 

 Let us now project the Hamiltonian on the qubit subspace, with the projector $\widehat \Pi_\mathrm{eg} = \ket{e}\!\bra{e}+\ket{g}\!\bra{g}$:
 \begin{align}
     \widehat H_\mathrm{m} & = \widehat \Pi_\mathrm{eg} \widehat H \widehat \Pi_\mathrm{eg} = \frac{\hbar \omega_\mathrm{ge}}{2} \hat \sigma_z + \hbar\Delta_\mathrm{R} \hat a^\dagger \hat a \nonumber\\
     & + \hbar g \left(c_0 + c_x \hat{\sigma}_{x}+ c_z \hat{\sigma}_{z} \right) \cdot \left( \alpha \hat a^\dagger + \alpha^* \hat a \right), 
 \end{align}
with 
\begin{align*}
c_0 &= \left(\bra{g} \hat\beta\ket{g} + \bra{e} \hat\beta\ket{e}\right)/2\\ c_x &= \bra{g} \hat\beta\ket{e}\\ c_z &= \left(\bra{e} \hat\beta\ket{e} - \bra{g} \hat\beta\ket{g}\right)/2,
\end{align*} 
where $\hat\beta \equiv \cos(\hat \varphi_{\mathrm{Q}}-\varphi_{\mathrm{ext}}).$
The interesting processes occur when the cavity drive is nearly resonant with one of the two sidebands, $\Delta_\mathrm{R} \sim \pm \omega_\mathrm{ge}$. We treat separately the two cases by going to the interaction picture with respect to $\widehat H_0^\pm = \hbar \omega_\mathrm{ge} \hat \sigma_z/2 \pm \hbar \omega_\mathrm{ge}\hat a^\dagger \hat a$:
\begin{align}
    \widehat H_\mathrm{m}^\pm = & \hbar \Delta_\mathrm{R}^\pm \hat a^\dagger \hat a \nonumber\\
    &+ \hbar g (c_0 + c_x (\hat \sigma^- e^{\mp i\omega_\mathrm{ge} t} + \hat \sigma^+ e^{\pm i\omega_\mathrm{eg} t})
    + c_z \hat \sigma_z) \nonumber\\
    &\cdot (\alpha \hat a^\dagger  e^{\pm i\omega_\mathrm{ge} t}+ \alpha^* \hat a  e^{\mp i\omega_\mathrm{ge} t}),
\end{align}
where $\widehat H_\mathrm{m}^+$ (respectively $\widehat H_\mathrm{m}^-$) is the Hamiltonian in the rotating frame $\widehat H_0^+$ (respectively $\widehat H_0^-$), and $\Delta_\mathrm{R}^\pm = \Delta_\mathrm{R} \pm \omega_\mathrm{ge}$ is the drive detuning with respect to the upper or lower sideband. 
Since the qubit-cavity system operates deep in the resolved sideband regime (at the bias point chosen for sideband preparation, $2 \omega_\mathrm{ge}/\kappa \approx 10$),  
we can safely neglect fast rotating terms, which yields:
\begin{equation}\label{eq:ASmy19}
    \hat H_\mathrm{m}^\pm \approx \hbar \Delta_\mathrm{R}^\pm \hat a^\dagger \hat a + \hbar g c_x (\alpha \hat \sigma^\pm \hat a^\dagger + \alpha^* \sigma^\mp \hat a).
\end{equation}

We proceed with the adiabatic elimination of the cavity field~\cite{Leghtas2015}, since the cavity dissipation $\kappa \mathcal{D}_{\hat a}(\hat \rho)$ dominates over the coupling $g c_x |\alpha |$.
We define the parameter $\epsilon \equiv g c_x |\alpha| /\kappa$,  with respect to which we can expand the density matrix $\hat \rho$, with $\hat \rho_{mn} = \bra{m}\hat \rho \ket{n}$ acting on the qubit's subspace:
\begin{align}\label{eq:lind}
    \frac{1}{\kappa}\frac{d \hat \rho }{dt} & =  -\frac{i}{\hbar\kappa} [\widehat H_\mathrm{m}^{\pm}, \hat \rho ] + \mathcal{D}_{\hat a}(\hat \rho), \\
    \hat \rho & =  \hat\rho_{00} \ket{0}\!\bra{0} + \epsilon \big(\hat\rho_{10} \ket{1}\!\bra{0} + \rho_{01} \ket{0}\!\bra{1} \big) + \nonumber\\
    & + \epsilon^2 \big(\hat\rho_{11} \ket{1}\!\bra{1} + \hat\rho_{02} \ket{0}\!\bra{2} + \hat\rho_{20} \ket{2}\!\bra{0}  \big) + O(\epsilon^3).\nonumber
\end{align}
The goal of the adiabatic elimination procedure is to obtain the reduced dynamics of the qubit alone
\begin{equation}
    \frac{d \hat{\rho}_\mathrm{Q}}{d t} = \mathrm{Tr}_\mathrm{R} \left[ \frac{d \hat{\rho}}{d t} \right]  = \frac{d \hat{\rho}_{00}}{d t} + \epsilon^2 \frac{d \hat{\rho}_{11}}{d t} + O(\epsilon^3),
\end{equation}
which can be done by projecting the Lindblad evolution of Eq.~\eqref{eq:lind} on the resonator's elements $\ket{0}\!\bra{0}, \ket{0}\!\bra{1}, \ket{1}\!\bra{1}$, so that
\begin{eqnarray}\label{eq:final}
    \frac{1}{\kappa}\frac{d \hat \rho_{00} }{d t} & = & -i \epsilon^2 \big( \hat \sigma^{\pm} \hat \rho_{10} - \hat \rho_{01} \hat \sigma^{\mp} \big) + \epsilon^2 \hat \rho_{11} + O(\epsilon^3)\label{eq:rho_00}\\
    \frac{1}{\kappa}\frac{d \hat \rho_{10}}{d t}  & = &  -i \hat \sigma^{\mp}\hat \rho_{00} - \Big( i \frac{\Delta_\mathrm{R}^\pm}{\kappa} -\frac{1}{2}\Big)\hat \rho_{10} + O(\epsilon) \label{eq:rho_10}\\
    \frac{1}{\kappa}\frac{d \hat \rho_{11}}{d t}  & =  & -i \big( \hat \sigma^{\mp} \hat \rho_{01} - \hat \rho_{10} \hat \sigma^{\pm} \big) - \hat \rho_{11} + O(\epsilon).\label{eq:rho_11}
\end{eqnarray}
Eq.~\eqref{eq:rho_00} shows that $\hat \rho_{00}$ is slowly varying, since its derivative is of order $O(\epsilon^2)$. In the rhs of Eq.~\eqref{eq:rho_10}, the first term is a source term, and the second one a damping term. As the source term is slowly varying, we can assume that $\hat \rho_{10}$ is always in its stationary state : $\frac{1}{\kappa}\frac{d \hat \rho_{10}}{d t}\approx 0.$ The same argument applies to Eq.~\eqref{eq:rho_11}, such that $\frac{1}{\kappa}\frac{d \hat \rho_{11}}{d t}\approx 0.$
In the end, we obtain\begin{eqnarray}
    \hat \rho_{10} & = & - \frac{\kappa}{\Delta_\mathrm{R}^\pm + i \frac{\kappa}{2}} \hat \sigma^{\mp} \hat \rho_{00} +O(\epsilon)\\
    \hat \rho_{11} & = & \frac{\kappa^2}{{\Delta_\mathrm{R}^\pm}^2 + \frac{\kappa^2}{4}} \hat \sigma^{\mp} \hat \rho_{00}  \hat \sigma^{\pm} + O(\epsilon)
\end{eqnarray}
By inserting these expressions in Eq.~\eqref{eq:final}, we recognize the Lindblad evolution associated to the following effective loss operators on the qubit:
\begin{equation}
    \widehat L^\pm = g c_x |\alpha| \sqrt{\frac{\kappa}{{\Delta_\mathrm{R}^\mp}^2 + \tfrac{\kappa^2}{4}}} \hat \sigma^{\pm}. 
\end{equation}
When the drive is set at resonance with one of the sidebands ($\Delta_\mathrm{R}^\pm = 0$), we retrieve Eq.~\eqref{eq:qubitDissipator}.

\newcommand{\sigmax}{\braket{\hat \sigma_x}}
\newcommand{\sigmay}{\braket{\hat \sigma_y}}
\newcommand{\sigmaz}{\braket{\hat \sigma_z}}

\begin{figure*}
    \centering
    \includegraphics[width=0.95\textwidth]{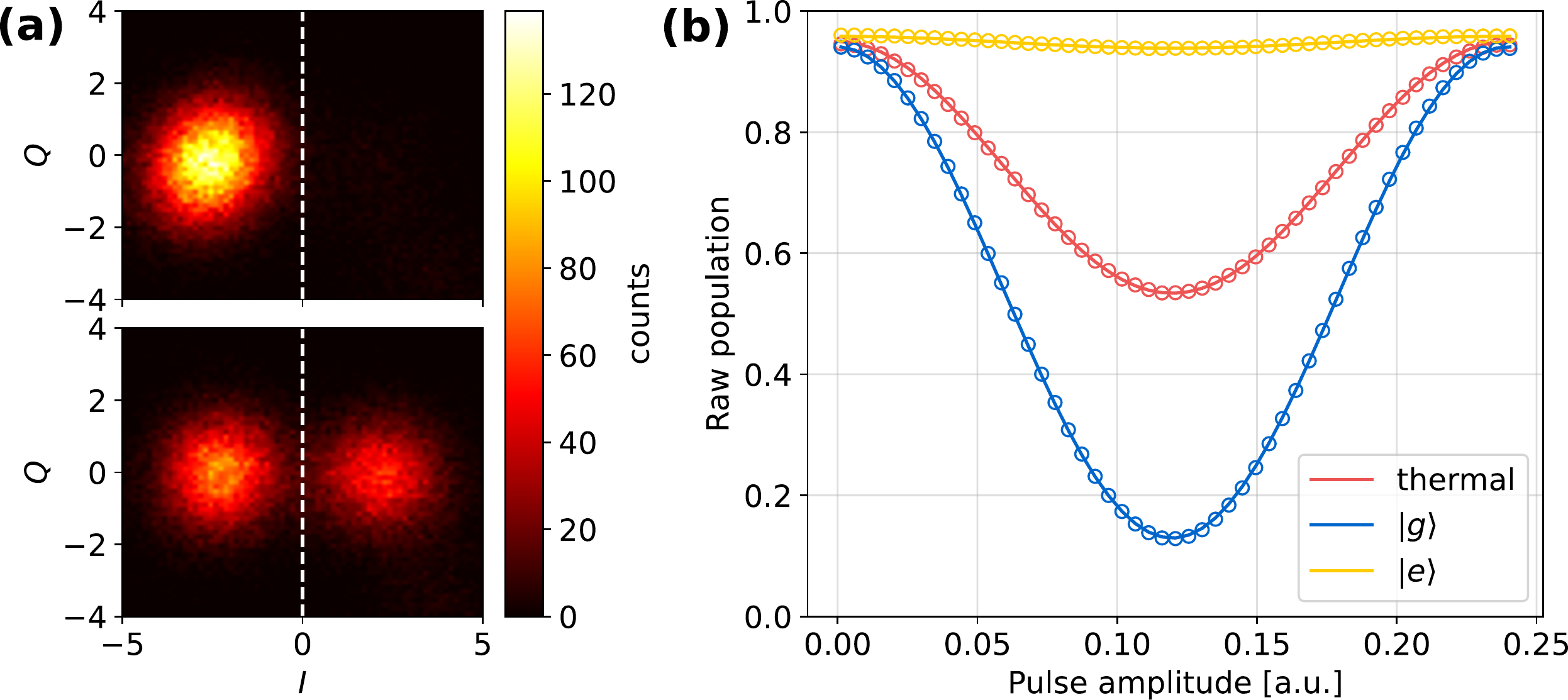}
    \label{fig:Fidelity_fitting}
    \caption{(a) Histograms of the 
    real (I) and imaginary (Q) parts of the cavity reflection coefficient, as obtained by demodulating and integrating a 600~ns pulse. 
    The x- and y-axis have been rescaled by the standard deviation of the Gaussian envelope. 
    Top plot is the histogram obtained when the system is initially prepared in its thermal state. 
    Even though $|g\rangle$ and $|e\rangle$ are almost equally populated, the dispersive is insufficient to separate the two distributions. 
    Bottom plot is the histogram obtained after a $|g\rangle \rightarrow |h\rangle \; \pi\mathrm{-pulse}$. 
    The left-blob corresponds to the $|e\rangle$ population, unaffected by the $\pi\mathrm{-pulse}$, while the right blob is due to the population transferred in $|h\rangle\rangle$. 
    Setting a threshold at $I = 0$ (white dashed line) implements a single shot readout. (b) Calibration of the state-preparation fidelity. 
    The qubit is prepared through sideband cooling in $|e\rangle$ (yellow), $|g\rangle$ (blue), or a thermal state (red). We apply a 64 ns pulse at the $|g\rangle$ - $|h\rangle$ transition frequency with a varying amplitude. 
    The dynamics is then fitted using Eq.~\eqref{eq:fidelity_equation} to extract the preparation fidelity (see main text for details).}
\end{figure*}

\section{Preparation fidelity estimation}
\label{app:fidelity}

To evaluate the preparation fidelity, we prepare the qubit in either $|g\rangle$, $|e\rangle$ or a thermal state. The density matrix reduced to the ge-manifold reads,
\begin{equation*}
    \rho_0^{\mathrm{prep}~j} = P_g^{\mathrm{prep}~j} |g\rangle\langle g| + (1-P_g^{\mathrm{prep}~j}) |e\rangle\langle e|,
\end{equation*}
where $P_g^{\mathrm{prep}~j}$ is the probability to be in the $|g\rangle$ state after preparation in state $ j \in \{\mathrm{g}, \mathrm{e}, \mathrm{th}\}$.  
Since the dispersive shift of the readout cavity is too small to directly distinguish $\ket{g}$ from $\ket{e}$, we apply a $64$~ns pulse, resonant with the $|g\rangle \rightarrow |h\rangle$ transition. 
The initial $|g\rangle$ population undergoes Rabi oscillations of angle $\theta$, while the $|e\rangle$ population remains unaffected.
\begin{equation}
\label{eq:rho_prep}
\begin{aligned}
    \rho^{{\mathrm{prep}~j}}(\theta) = & P_g^{\mathrm{prep}~j}\frac{1+\cos{\theta}}{2}|g\rangle \langle g| \\
    & + P_g^{\mathrm{prep}~j}\frac{1-\cos{\theta}}{2}|h\rangle \langle h| \\
    & + (1-P_g^{\mathrm{prep}~j})|e\rangle\langle e|\\
    & + P_g^{\mathrm{prep}~j}\frac{|\sin{\theta}|}{2}(|g\rangle \langle h| + |h\rangle \langle g|).
\end{aligned}
\end{equation}
Decoherence has been neglected in this process as the pulse duration is short compared to the decoherence rates of the $|g\rangle$ - $|h\rangle$ transition. 
We then read out the state of the qubit through the dispersive shift of the readout resonator. 
Histograms of the real (I) and imaginary part (Q) of the reflection coefficient, measured with a $600$~ns pulse are plotted in Fig.~\ref{fig:Fidelity_fitting}a. 
The continuous variable $I$ is then compared to a threshold $I_\mathrm{threshold}=0$ to yield a Boolean detection result $\mathrm{left} = I<I_\mathrm{threshold}$. 
By averaging a large number of repetitions, we measured the probability $P^{\mathrm{prep}~j}[\mathrm{left}](\theta)$ to obtain $I<I_\mathrm{threshold}$ after a Rabi pulse of angle $\theta$ (see Fig.~\ref{fig:Fidelity_fitting}b), for each preparation protocol. 
This probability is given by:
\begin{equation}
P^{\mathrm{prep}~j}[\mathrm{left}](\theta) = \sum_{x\in\{g, e, h\}} P[\mathrm{left}|x]\langle x|\rho^{\mathrm{prep}~j}(\theta)|x\rangle,
\end{equation} 
where $P[\mathrm{left}|g]$,  $P[\mathrm{left}|e]$ and $P[\mathrm{left}|h]$ are the conditional probabilities of measuring $I<I_\mathrm{threshold}$ knowing that the qubit was in the state $|g\rangle$ $|e\rangle$ or $|h\rangle$ respectively. 
In a perfect detection scenario, $P[\mathrm{left}|g] = P[\mathrm{left}|e] = 1$ and $P[\mathrm{left}|h] = 0$.  
Combining Eq.~\eqref{eq:rho_prep} and Eq.~\eqref{eq:fidelity_equation}, we arrive at
\begin{equation}
    \label{eq:fidelity_equation}
    \begin{aligned}
    P^{\mathrm{prep}~j}[\mathrm{left}](\theta) = &P_g^{\mathrm{prep}~j}\frac{P[\mathrm{left}|g]+P[\mathrm{left}|h]}{2}\\ 
    & + (1-P_g^{\mathrm{prep}~j})P[\mathrm{left}|e]\\
    & + P_g^{\mathrm{prep}~j} \cos(\theta) \frac{P[\mathrm{left}|g]-P[\mathrm{left}|h]}{2}.   
    \end{aligned}
\end{equation}
Considering the large occupation of the thermal bath, we assume equal populations in $\ket{g}$ and $\ket{e}$: $P_{g}^{\mathrm{prep~th}} = 0.5$. 
We proceed by fitting the 3 curves of Fig.~ \ref{fig:Fidelity_fitting} with the free parameters $P[\mathrm{left}|g]$, $P[\mathrm{left}|e]$, $P[\mathrm{left}|h]$, $P_g^\mathrm{prep~g}$ and $P_g^\mathrm{prep~e}$. We obtain the conditional readout probabilities $P[\mathrm{left}|g] = 94.04 \pm 0.04\,$\%, $P[\mathrm{left}|e] = 95.87 \pm 0.03\,$\% and $P[\mathrm{left}|h] = 10.99 \pm 0.05\,$\%. 
The values of $1 - P[\mathrm{left}|g] = 6~$\%, $1 - P[\mathrm{left}|e] = 4~\%$ correspond to the mislabeling, due to the overlap of the Gaussian distributions. 
The larger value of $P[\mathrm{left}|h] = 11\,\%$ is due to the decay of the $|h\rangle$ state (of lifetime 7 $\mu$s) during the 600~ns readout pulse. 
The extracted preparation fidelity for the $|g\rangle$ and $|e\rangle$ states are $P_g^{\mathrm{prep~g}} = 97.67\pm 0.05\,$ \% and $1-P_g^{\mathrm{prep~e}} = 97.69\pm 0.08\,$ \%. 
The quoted error intervals are obtained by a bootstrap technique: the fit is repeated on a subset of the data obtained by random sampling with replacement of the data point, from which, the mean value and standard deviation of each parameter is extracted.  
The effective temperature of the qubit after the preparation is,
\begin{equation*}
    T_g^\mathrm{prep~g} = \frac{\hbar \omega_\mathrm{ge}}{k_B \left[\log(P_g^{\mathrm{prep~g}}) - \log (1-P_g^{\mathrm{prep~g}})\right]} = 23\,\mu\mathrm{K}.
\end{equation*}

\begin{figure*}[t]
    \centering
    \includegraphics[width=0.95\textwidth]{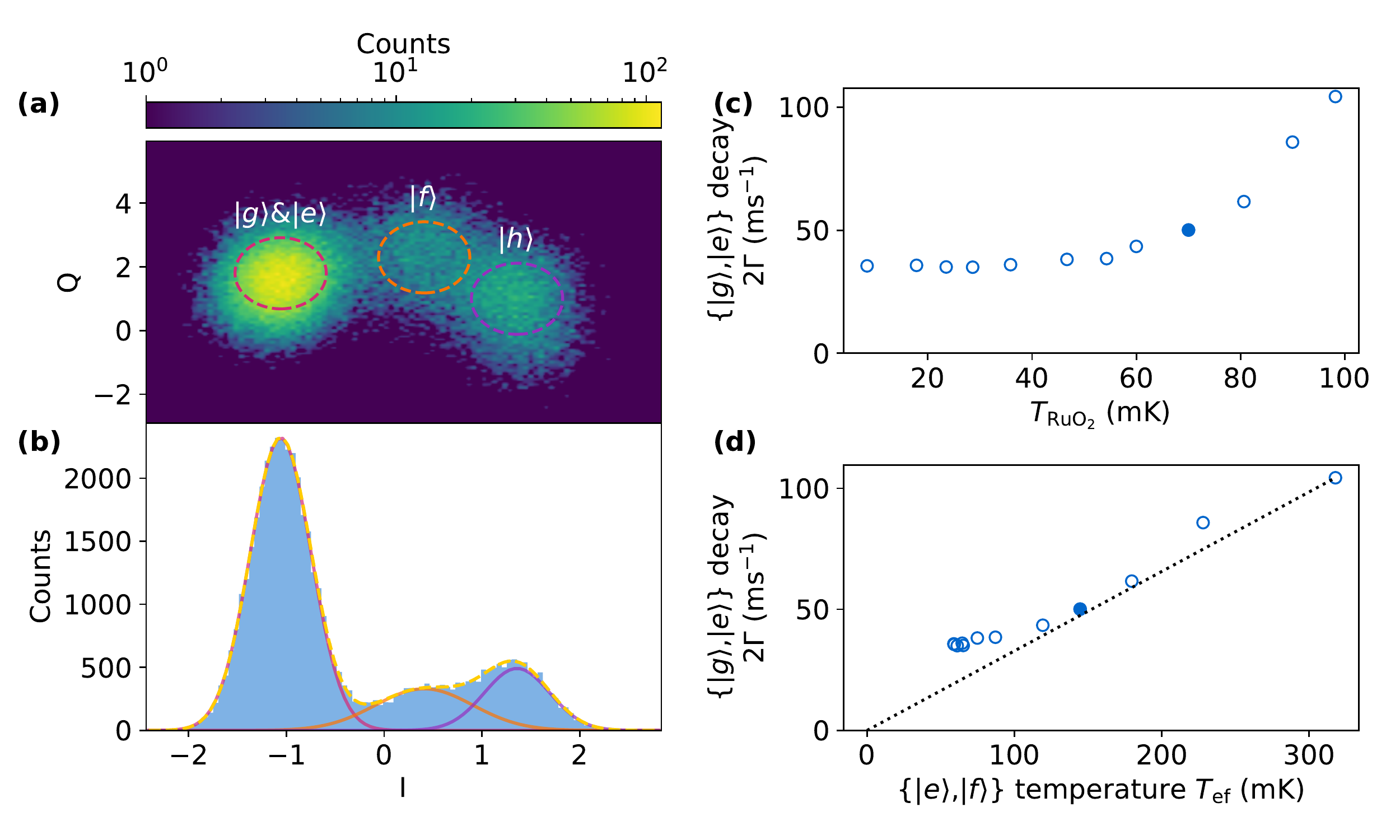}
    \label{fig:Temperature_dependency}
    \caption{Temperature dependence of the decoherence rate. (a) 2-dimensional histogram of the I, Q quadratures of the readout reflection coefficients for a qubit at thermal equilibrium with the environment. (b) Histogram of the I quadrature. The population in the manifolds $\{\ket{g}, \ket{e}\}$, $\ket{f}$ and $\ket{h}\}$ are determined by fitting the various peaks with Gaussian functions (see text for details). 
    (c) Energy decay rate $2 \Gamma$ in the $\{\ket{g},\ket{e}\}$ manifold measured via $T_1$ relaxometry (see Fig.~\ref{fig:coherence}a), as a function of cryostat temperature. 
    (d) The same data are plotted as a function of the effective temperature of the $\ket{e} \rightarrow\ket{f}$ transition, as determined from Eq.~\eqref{eq:temp_fit_equation}. The filled point is the one extracted from the histograms in (a) and (b). 
    The dashed line is a guide to the eyes highlighting the linear dependency above $T_\mathrm{ef}=100$~mK.}
\end{figure*}
\section{Noise temperature of the qubit environment}
\label{app:t1_vs_t}
In this section, we determine how the decoherence rate varies with temperature. 
To achieve this, we heat the mixing-chamber of the cryostat with a resistor. 
The temperature $T_{\mathrm{RuO}_2}$, as measured by a Ruthenium oxide probe built-in with the cryostat (model Bluefors BF-LD250) is stabilized thanks to a feedback loop to various setpoints ranging from 7~mK to 100~mK.  
For each point, we measure the decay rate $2 \Gamma$ of the states $\ket{e}$ and $\ket{g}$, akin to the measurement presented in the main text (refer to Fig.~\ref{fig:coherence}a).
We observe a nearly constant decay rate $2 \Gamma \approx 35~\mathrm{ms}^{-1}$ in the range $7$~mK$\le T_{\mathrm{RuO}_2} \le 50$~mK. 
Above 50~mK, we observe a linear increase of the decay rate, compatible with an imperfect thermalization of the sample with the mixing chamber. 
The fact that the asymptote of the curve $2 \Gamma (T_{\mathrm{RuO}_2})$ doesn't intersect with the origin is attributed to a possible miscalibration of the cryostat temperature sensor at high temperature. 

In order to obtain an independent temperature measurement, we use the residual thermal population of the higher qubit excited states $\{\ket{f}, \ket{h}\}$. 
This signal serves as a local probe, scrutinizing the noise temperature of the circuit at the second transition frequency of 3.7~GHz. 
In practice, we let the circuit thermalize with its environment, and then record a histogram of the real ($I$) and imaginary part ($Q$) of the readout cavity reflection coefficient, as visible on Fig.~ \ref{fig:Temperature_dependency}b. 
Three peaks are visible on the histogram, corresponding to the population in the manifold $\{\ket{g}, \ket{e}\}$, the state $\ket{f}$ and the state $\ket{h}$ respectively. 
We assume a Boltzmann distribution for the population in the various qubit states: $p_k \propto e^{-k_b T_\mathrm{eff}/E_k}$, where $E_k$ is the energy of state $k$ ($k \in \{\ket{g}, \ket{e}, \ket{f}, \ket{h}\}$. 
Furthermore, by neglecting the small transition frequencies $ \omega_\mathrm{ge} / 2 \pi \sim \qubitfreq$~MHz, 
and $\omega_\mathrm{fh}/2 \pi \sim 50$~MHz, compared to $\omega_\mathrm{ef}/2\pi \sim 3.7$~GHz, we get $p_{\ket{g}} = p_{\ket{e}} \equiv p_\mathrm{g,e}/2$ and $p_{\ket{f}} = p_{\ket{h}} \equiv p_\mathrm{f,h}/2$. 
We extract the populations $p_\mathrm{g,e}$ and $p_\mathrm{f, h}$ by a triple Gaussian fit to the readout histogram, where the Gaussian peaks corresponding to $\ket{f}$ and $\ket{h}$ are constrained to the same area. 
From the values $p_\mathrm{g,e}$ and $p_\mathrm{f, h}$, we determine the effective temperature:
\begin{equation}
    \label{eq:temp_fit_equation}
    T_\mathrm{ef} = \frac{\hbar \omega_\mathrm{ef}}{k_B \log \left(\frac{p_\mathrm{g,e}}{p_{f,h}}\right)}.
\end{equation}

We then plot the decay rate $2 \Gamma$ as a function of effective temperature 
$T_\mathrm{ef}$ in Fig.~\ref{fig:Temperature_dependency}d. 
We observe a linear dependence on most of the temperature range indicating that the $\ket{g} \rightarrow \ket{e}$ and $\ket{e} \rightarrow \ket{f}$ transitions are coupled to thermal environments with similar noise temperatures, in spite of their 3-orders of magnitude frequency difference.

\section{Charge spectrum analyzer}
\label{app:spectrum}
In this section, we develop a theoretical model for the expected signal-to-noise ratio in the frequency-resolved charge detection experiment.

\subsection{Qubit evolution during the interrogation time}
We first model the evolution of the qubit during the interrogation time, by taking into account the interaction with the calibration-tone (Rabi-frequency $\Omega_r$, finite detuning of the calibration tone $\Delta = \omega_\mathrm{ge} - \omega_\mathrm{cal}$. Since the qubit is coupled to a thermal bath with a large occupation, we choose an equal rate $\Gamma$ for the loss and gain of qubit excitations. From the empirical finding $T_1 \approx T_2$ (see Fig.~\ref{fig:coherence}), we also assume a dephasing rate $\Gamma_\phi \approx \Gamma/2$. 
The full evolution of the qubit's density matrix $\rho$ is thus, in a frame rotating at the drive frequency: 
\begin{align}
\frac{d \hat \rho}{d t} = -\frac{i}{\hbar} [\hat H, \hat \rho] &+ \Gamma \left[ \hat \sigma \rho \hat \sigma^\dagger - \frac{1}{2} (\hat \sigma^\dagger \hat \sigma \hat \rho + \hat \rho \hat \sigma^\dagger \hat \sigma )\right] \\
 &+  \Gamma \left[ \hat \sigma^\dagger \rho \hat \sigma - \frac{1}{2} (\hat \sigma \hat \sigma^\dagger \hat \rho + \hat \rho \hat \sigma \hat \sigma^\dagger )\right] \nonumber\\
  &+  \frac{\Gamma}{2} \left[ \hat \sigma_z^\dagger \rho \hat \sigma_z - \frac{1}{2} (\hat \sigma_z \hat \sigma_z^\dagger \hat \rho + \hat \rho \hat \sigma_z \hat \sigma_z^\dagger )\right],\nonumber 
\end{align}
with $\hat H = \hbar \Delta \hat \sigma_z/2 + \hbar \Omega_\mathrm{r} \hat \sigma_x/2$. 
We proceed by calculating the Bloch equations for the 3 components of the qubit pseudo-spin:
\begin{align}
    \frac{d \sigmax}{dt} &= - \Delta \sigmay - 2 \Gamma \sigmax \\
    \frac{d \sigmay}{dt} &= - \Omega_\mathrm{r} \sigmaz + \Delta \sigmax - 2 \Gamma \sigmay \\
    \frac{d \sigmaz}{dt} &= \Omega_\mathrm{r} \sigmay  - 
 2 \Gamma \sigmaz. 
\end{align}
These equations describe a rotation around an axis $\Delta \mathbf{e_z} + \Omega_\mathrm{r} \mathbf{e_x}$ combined with an isotropic relaxation towards the origin of the Bloch sphere at a rate $\Gamma$ due to the various relaxation channels. 
We solve for a qubit initially prepared in $\ket{g}$ ($\sigmaz = -1, \sigmax = \sigmay = 0$), and obtain:
\begin{align*}
    \sigmax &= e^{-2\Gamma t} \left(\cos(\sqrt{\Omega_\mathrm{r}^2 + \Delta^2} t) - 1 \right) \frac{\Delta \Omega_\mathrm{r}}{\Omega_\mathrm{r}^2 + \Delta^2} \\
    \sigmay &= e^{-2\Gamma t} \sin(\sqrt{\Omega_\mathrm{r}^2 + \Delta^2} t) \frac{\Omega_\mathrm{r}}{\sqrt{\Omega_\mathrm{r}^2 + \Delta^2}} \\
    \sigmaz &= - e^{-2\Gamma t} \left(\Delta^2 + \Omega_\mathrm{r} ^2 \cos(\sqrt{\Omega_\mathrm{r}^2 + \Delta^2} t) \right) \frac{1}{\Omega_\mathrm{r}^2 + \Delta^2}. \\
\end{align*}
At the end of the interrogation time, we can thus obtain the magnitude of the pseudo-spin projection in the $x, y$ plane:
\begin{equation}
    \label{eq:sigma0}
    |2 \braket{\hat \sigma}_0| = |\sigmax + i \sigmay| = \Omega_\mathrm{r} \tau_I  e^{-2 \Gamma \tau_I} f(\Delta),
\end{equation}
where $f(\Delta)$ is the frequency response function of the detector, given by
\begin{align}
    f(\Delta) &= \sqrt{\frac{\Delta^2 \mathrm{sinc}^2(\sqrt{\Omega_r^2 + \Delta^2} \frac{\tau_I}{2 \pi}) + \Omega_\mathrm{r}^2 \mathrm{sinc}^2(\sqrt{\Omega_r^2 + \Delta^2} \frac{\tau_I}{\pi})}{\Omega_\mathrm{r}^2 + \Delta^2}} \nonumber\\
    &\approx \mathrm{sinc}\left(\frac{\Delta}{\Omega_\mathrm{full}}\right).
    \label{eq:bandwidth}
\end{align}
Where $\Omega_\mathrm{full} = \frac{2 \pi}{\tau_I}$ and the convention $\mathrm{sinc}(x) = \sin(\pi x)/\pi x$ has been used. The second equation is valid in the limit $\Omega_\mathrm{r} \tau_I \ll 1$.

\subsection{Signal processing} 
At the end of the interrogation time, a projective measurement of one of the transverse components of the pseudo-spin is performed in the qubit frame. 
\begin{equation}
    \braket{m_k} = 1/2 + \Re\left[\braket{\sigma}_0 (-i)^k e^{i \Delta k \tau}\right].
\end{equation}
The term $(-i)^k$ encodes for the alternating measurement basis $\{X, Y, -X, -Y\}$. The term $e^{i \Delta k \tau}$ describes the phase difference between the frames of the qubit and calibration tone. Without loss of generality, we can ignore the phase of $\braket{\sigma}_0$ and assume $\braket{\sigma}_0 \in \mathds{R}^+$, such that
\begin{equation}
    \braket{m_k} = 1/2 + \braket{\sigma}_0 \Re\left[ (-i)^k e^{i \Delta k \tau}\right].
\end{equation}
Finally, samples undergo the transformation $\sigma_k = i^k (m_k - 1/2)$. We thus get 
\begin{align}
    \label{eq:sigma}
    \braket{\sigma_k} &= \left\{
        \begin{array}{ll}
            \braket{\sigma}_0 \cos(\Delta k \tau) & \quad k \mathrm{\,\,even} \\
            i \braket{\sigma}_0 \sin(\Delta k \tau) & \quad k \mathrm{\,\,odd}.
        \end{array}
    \right.
\end{align}
Hence, the real and imaginary parts of the complex values $\braket{\sigma}_0 e^{i \Delta t}$ are encoded pairwise on the successive samples $\sigma_k$. 
The records are then grouped by windows of $N = 1000$ consecutive samples, and Fourier transformed to yield periodograms. In order to reduce the spacing between adjacent frequency bins, we perform the Fourier transform on a 0-padded version of the samples $\{z_k\}_{0\le k < N_p N}$, with
\begin{equation}
    \label{eq:zk}
    z_k = \left\{
        \begin{array}{ll}
            \sigma_k & \quad 0 \le k \le N-1 \\
            0 & \quad N \le k < N_p N.
        \end{array}
    \right.
\end{equation}
The padding factor $N_p$ represents the number of frequency bins in each measurement bandwidths. We typically use $N_p = 5$ in our data analysis. 
We denote $\{Z_k\}_{0 \leq k < N_p N}$ the Fourier transform of the samples $\{z_n\}_{0 \leq n < N_p N}$:
\begin{equation}
    \label{eq:fft}
    Z_n = \sum_{k=0}^{N_p N-1} z_k e^{-2 i \pi k n/N_p N}.
\end{equation}
Following Bartlett's method, the spectrum is then estimated by taking the mean-value $S_n = \langle |Z_n|^2 \rangle$ over a large number of periodograms. 

\subsection{Response to the calibration tone and frequency aliasing}

\begin{figure}    \includegraphics[width=0.48\textwidth]{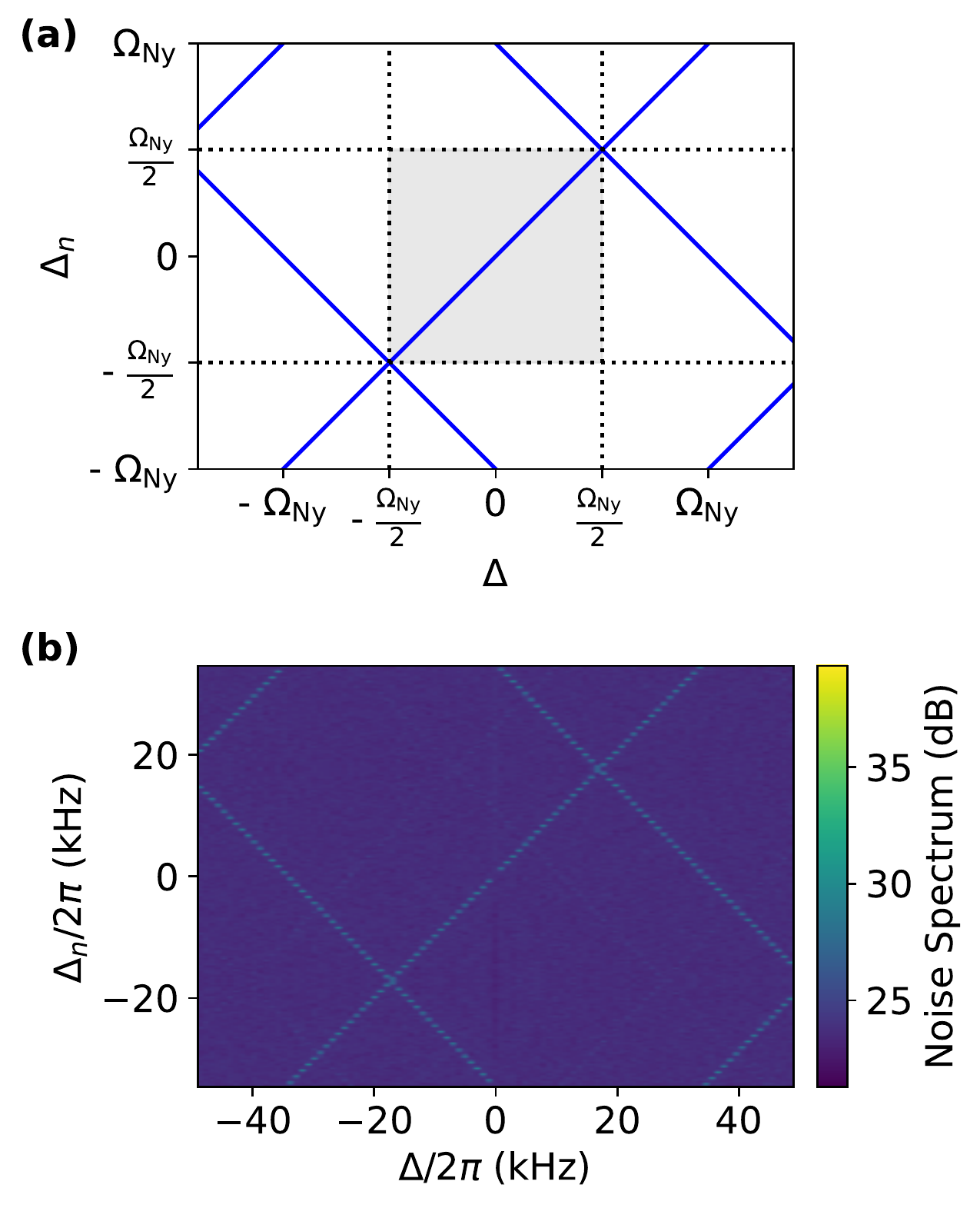}
    \caption{Aliasing diagram of the charge spectrum analyzer: (a) The blue lines indicate the position of the peaks in discrete frequency space $\Delta_n$ as a function of the continuous frequency $\Delta$ of the applied tone, $\Delta \equiv \Delta_n \;(\bmod\; 2 \Omega_\mathrm{Ny})$ and $\Delta \equiv \Omega_\mathrm{Ny} - \Delta_n \;(\bmod\; 2 \Omega_\mathrm{Ny})$ (see main text for details). (b) Experimental spectrogram with an applied tone of fixed frequency. The detuning $\Delta$ is swept by adjusting the qubit local oscillator frequency. In this particular instance, the repetition time of the experiment was set to $\tau = 14.4~\mu$s, corresponding to a Nyquist frequency $\Omega_\mathrm{Ny}/2\pi = 35~$kHz. The frequency axis in the Figure \ref{fig:spectrum}c has been cropped to only display frequencies from $-\Omega_\mathrm{Ny}/2 \le \Delta_n \le \Omega_\mathrm{Ny}/2$.}
    \label{fig:aliasing}
\end{figure}

Because of the calibration tone, the samples $z_k$ have a non-zero expectation value (see Eq.~\eqref{eq:sigma}). We now estimate the lineshape $S^\mathrm{signal}_n = |\braket{Z_n}|^2$ resulting from this signal. 
By combining Eq.~\eqref{eq:sigma} with Eq.~\eqref{eq:fft}, and separating the contribution of even and odd index $k$ in the sum, we get:
\begin{align*}
    \braket{Z_n} =& \braket{\sigma}_0 \sum_{k=0}^{N/2 - 1} \cos(2 \pi \frac{2 k \Delta}{2 \Omega_\mathrm{Ny}}) e^{-i 2 \pi \frac{2 k \Delta_n}{2 \Omega_\mathrm{Ny}}} \\ 
    &+ i \sin( 2 \pi \frac{(2k + 1) \Delta}{2 \Omega_\mathrm{Ny}}) e^{-i  2 \pi \frac{(2 k+1) \Delta_n}{2 \Omega_\mathrm{Ny}}},
\end{align*}
where the $n^\mathrm{th}$ frequency bin is given by $\Delta_n = \frac{2 \pi n}{\tau N_p N}$ and the Nyquist frequency $\Omega_\mathrm{Ny} = \pi/\tau$. 
After elementary arithmetic manipulations, we arrive at:
\begin{align*}
    \braket{Z_n} =& \frac{\braket{\sigma}_0}{2} \left[ e^{i 2 \pi \frac{\Delta - \Delta_n}{\Omega_\mathrm{Ny}} \frac{N-1}{4}} \frac{\sin(2 \pi\frac{\Delta - \Delta_n}{\Omega_\mathrm{Ny}} \frac{N}{4})}{\sin(2 \pi \frac{\Delta -\Delta_n}{4\Omega_\mathrm{Ny}})} \right. \\ 
    &+ \left. i e^{-i 2 \pi \frac{\Delta + \Delta_n}{\Omega_\mathrm{Ny}} \frac{N-1}{4}} \frac{\sin(2 \pi \frac{\Delta + \Delta_n}{\Omega_\mathrm{Ny}} \frac{N}{4})}{\cos(2 \pi \frac{\Delta +\Delta_n}{4\Omega_\mathrm{Ny}})} \right].
\end{align*}
For large values of $N$, we have
\begin{align*}
    \label{eq:freq_aliasing}
    S^\mathrm{signal}_n \approx &\frac{\braket{\sigma}_0^2}{4} \left[ \left(\frac{ \sin(2 \pi \frac{\Delta - \Delta_n}{\Omega_\mathrm{Ny}} \frac{N}{4})}{\sin(2 \pi \frac{\Delta - \Delta_n}{4 \Omega_\mathrm{Ny}})}\right)^2 \right.  \\ 
    &+ \left. \left(\frac{ \sin(2 \pi \frac{\Delta + \Delta_n}{\Omega_\mathrm{Ny}} \frac{N}{4})}{\cos(2 \pi \frac{\Delta + \Delta_n}{4 \Omega_\mathrm{Ny}})}\right)^2\right].
\end{align*}
This expression is peaked around the values $\Delta \equiv \Delta_n \;(\bmod\; 2 \Omega_\mathrm{Ny})$ and $\Delta \equiv \Omega_\mathrm{Ny} - \Delta_n \;(\bmod\; 2 \Omega_\mathrm{Ny})$. Fig.~\ref{fig:aliasing} shows the two families of peaks in the $(\Delta, \Delta_n)$ plane. In the aliasing-free region ${-\Omega_\mathrm{Ny}/2 \le \Delta \le \Omega_\mathrm{Ny/2}}$ highlighted by the grey square, the signal is given in a good approximation by
\begin{equation}
    \label{eq:signal}
    S^\mathrm{signal}_n \approx \left( \braket{\sigma}_0 \frac{N}{2} \mathrm{sinc}\left( \frac{\Delta - \Delta_n}{\Omega_\mathrm{RBW}}\right)\right)^2.
\end{equation}
In this expression, we use the definition $\mathrm{sinc}(x) = \sin(\pi x)/\pi x$ and the residual bandwidth of the measurement is given by
\begin{equation}
    \label{eq:rbw}
    \Omega_\mathrm{RBW} = 2 \pi / \tau N.
\end{equation}

\subsection{Signal-to-noise ratio}
Owing to the quantum nature of our sensor, the measurement records are in essence discrete, such that a fundamental sampling noise, of spectral shape $S^\mathrm{sampling}_n$, affects our measurement. Indeed, the spectrum estimator can be decomposed according to

\begin{equation}
    S_n = S^\mathrm{signal}_n + S^\mathrm{sampling}_n,
\end{equation}
with $S^\mathrm{signal}_n = |\braket{Z_n}|^2$ and $S^\mathrm{sampling}_n = \braket{|Z_n - \braket{Z_n}|^2}$. To calculate the sampling noise, we can consider the situation where no calibration tone is applied, such that the samples $\{ z_k \}_{0 \le k < N}$ are independent, with $\braket{z_k} = 0$ and $\braket{z_k z_k'^*} = \delta_{k,k'}/4$. Combined with the relation \eqref{eq:fft}, we get
\begin{equation}
    \label{eq:sampling_noise}
    S^\mathrm{sampling}_n = \langle | Z_k |^2\rangle = \sum_{k=0}^{N-1} \braket{z_k z_k'} = N/4.
\end{equation}
By combining the relations \eqref{eq:signal} and \eqref{eq:sampling_noise}, we get the signal-to-noise ratio:
\begin{equation}
    \label{eq:SNR}
    \mathrm{SNR} = \sqrt{\mathrm{max}(S^\mathrm{signal}_n) / S^\mathrm{sampling}_n} = \sqrt{N} \mathrm{\sigma}_0.
\end{equation}
The blue curve in the inset of Fig.~\ref{fig:spectrum}c is calculated using Eq.~\eqref{eq:SNR} and Eq.~\eqref{eq:sigma0}, with $2 \Gamma = (24~ \mu\mathrm{s})^{-1}$, in qualitative agreement with the value obtained with a more direct measurement (see main text), and a 84 \%  scaling factor to account for finite readout efficiency.

\subsection{Approximate expression for the optimal charge sensitivity}

The noise spectrum in units of $\mathrm{e}^2/\mathrm{Hz}$ is calibrated such that the area under the calibration peak matches the known modulation amplitude:
\begin{equation}
    \label{eq:area}
    \int_{0}^\infty S_{ee} d\omega/2\pi = (2 N_\mathrm{drive})^2,
\end{equation}
where the factor 2 accounts for the number of elementary charges in each Cooper-pair. 
The left-hand side of Eq.~\eqref{eq:area} is approximately given by $S_\mathrm{ee}[\omega_d] \cdot \Omega_\mathrm{RBW}/2\pi$, such that the peak of the noise spectrum is given by:
\begin{equation}
    \mathrm{max}(S_\mathrm{ee}^\mathrm{signal})= (2 N_\mathrm{drive})^2 \frac{2\pi}{\Omega_\mathrm{RBW}}.
\end{equation}
We can now use the definition of the signal-to-noise ratio (in conjunction with the linear relationship between $S_\mathrm{ee}$ and $S_n$):
\begin{equation}
\label{eq:See_signal}
    S_\mathrm{ee}^\mathrm{sampling} = \mathrm{max}(S_\mathrm{ee}^\mathrm{signal})/(\mathrm{SNR})^2.
\end{equation}
Additionally, by combining Eq.~\eqref{eq:sigma0} with Eq.~\eqref{eq:SNR}, we obtain the approximate expression of the signal-to-noise ratio for a calibration tone well within the detector bandwidth ($\Delta\ll\Omega_\mathrm{full})$:
\begin{equation}
    \label{eq:SNRapprox}
    (\mathrm{SNR})^2 = N  \left(\frac{\mathrm{exp}(-\tau_I/T_1) \Omega_R \tau_I}{2}\right)^2.
\end{equation}
\begin{figure}[t]   \includegraphics[width=0.48\textwidth]{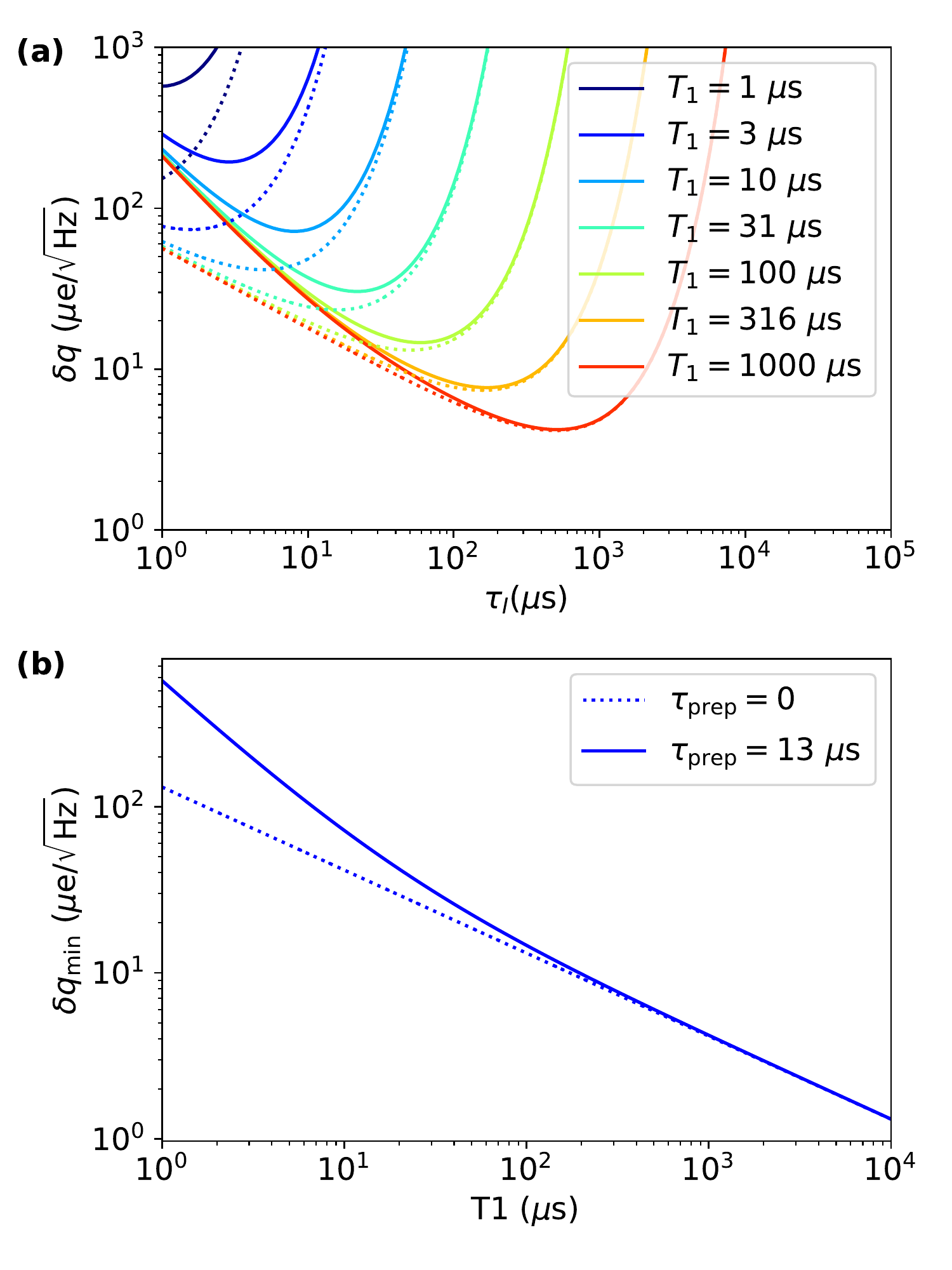}
    \caption{Sensitivity of the heavy-fluxonium charge spectrum analyzer. (a) Charge sensitivity as a function of interrogation time $\tau_I$, for various qubit lifetimes $T_1$ (see legend). The dashed line correspond to an ideal scenario where the duty-cycle $\tau_I / \tau = 1$ (see Eq. \eqref{eq:ideal_sensitivity}). The full line corresponds to a fixed preparation/readout time $\tau_\mathrm{prep} = 13~\mu$s. (b) Minimal sensitivity (obtained for an optimal value of $\tau_I$) as a function of $T_1$. The 2 scenarios considered in (a) are still represented by dashed and full lines respectively.}
    \label{fig:sensitivity}
\end{figure}
Finally, by inserting Eq.~\eqref{eq:SNRapprox} into Eq.~\eqref{eq:See_signal}, and using the expressions \eqref{eq:rabi} for $\Omega_r$ and \eqref{eq:rbw} for $\Omega_\mathrm{RBW}$, we derive 
\begin{equation}
    \label{eq:final_sensitivity}
    S_\mathrm{ee}^\mathrm{sampling} = \delta q^2 = \frac{4 \tau }{\tau_I^2 \omega_\mathrm{ge}^2 \pi^2 \exp{(-2 \tau_I/T_1)}}.
\end{equation} 

To minimize $\delta q$, it is beneficial to maximize the duty cycle $\tau_I/\tau$. 
In our experiment, the total preparation and readout time is approximately amounts to  $\tau_\mathrm{prep} \sim 13~\mu$s, rendering the total cycle time as $\tau = \tau_I + \tau_\mathrm{prep}$. 
Fig.~\ref{fig:sensitivity} illustrates the evolution of $S_\mathrm{ee}^\mathrm{sampling}$ as a function of the interrogation time $\tau_I$ for various $T_1$ values under two distinct scenarios. 

In the first one (dotted lines), we have considered the ideal case $\tau_\mathrm{prep} = 0~\mu$s. 
In this ideal case, the optimal sensitivity is obtained for $\tau_I = T_1/2$, reaching a value 
\begin{equation}
\label{eq:ideal_sensitivity}
\delta q_\mathrm{min}^2 = \frac{8 \exp(1)}{T_1 \omega_\mathrm{ge}^2 \pi^2}.    
\end{equation}
Remarkably, $\delta q_\mathrm{min}$ only depends on the qubit frequency $\omega_\mathrm{ge}$ and coherence time $T_1$. 
This stems from the observation that at the flux-frustration point, the Rabi frequency depends only on the product $\omega_\mathrm{ge} N_\mathrm{drive}$ (see Eq.~\eqref{eq:rabi}), and not on the specific qubit parameters, as long as the systems operates in the heavy-fluxonium regime. 

In the second scenario (full lines in Fig. \ref{fig:sensitivity}), we consider a realistic preparation and readout time $\tau_\mathrm{prep} = 13~\mu$s. 
As evident from Eq.\ref{eq:final_sensitivity}, for a given interrogation time $\tau_I$, the sensitivity is degraded by a factor $\sqrt{1/\eta}$, where $\eta = \tau_I/\tau$ denotes the duty cycle of the experiment, in comparison to the ideal case.
However, the optimal sensitivity, defined as
\begin{equation}
    \delta q_\mathrm{min} = \displaystyle \min_{\tau_I}(\delta q)
\end{equation}
remains close to the ideal one as long as  $\tau_\mathrm{prep} \ll T_1$, as visible in Fig.~\ref{fig:sensitivity}b.

\section{Estimate of the charge modulation by a DC-biased membrane}
\label{app:membrane}

In this section, we assess the possibility for the heavy-fluxonium to reach the strong-coupling regime with state-of-the art macroscopic electromechanical systems. 
For this, we estimate the magnitude of the charge modulation induced by the zero-point fluctuations of a DC-biased vacuum-gap capacitor. 
In this scenario, we consider that the out-of-plane vibrations of a silicone-nitride membrane modulate the capacitance between two parallel electrodes subjected to a DC bias voltage $V_g$.

\begin{table}
\begin{tabular}{|c|c|c|}
\hline
membrane side & $l$ & 150~$\mu$m\\
\hline
membrane stress & $\sigma$ & 1~GPa\\
\hline
mechanical mode frequency & $\Omega_m/2\pi$ & 1.8~MHz\\
\hline
motional mass & m & 3~ng\\
\hline
zero-point fluctuations & $x_\mathrm{zpf}$ & 7~fm\\
\hline
silicon nitride density & $\rho$ & 3200~kg.m$^{-3}$\\
\hline
capacitor electrodes distance & $h$ & 500~nm\\
\hline
electrode surface & $S$ & (90~$\mu$m)$^2$ \\
\hline
\end{tabular}
\caption{\label{table:membrane} Estimated parameters for a macroscopic electromechanical system.}
\end{table}

Table~\ref{table:membrane} summarizes the main geometric parameters of the membrane. The membrane lateral dimensions are chosen such that the fundamental mechanical mode matches the qubit frequency $\omega_\mathrm{ge}$~\cite{Yu2012}. 
The area of the electrodes are chosen to obtain a capacitance $C = 50~\mathrm{fF}$ matching the value reported in our fluxonium implementation. We assume an electrode separation $h= 500$~nm, which is a conservative estimate based on flip-chip assemblies already reported in the literature~\cite{Seis2022}.
The mechanical resonator undergoes the sum of the restoring force and the electrostatic force \begin{equation}
F=-m\Omega_m^2(z-h)-V_g^2\epsilon_0 S/2z^2.
\end{equation}
Mechanical stability requires $V_g<\sqrt{m\Omega_m^2h^3/\epsilon_0 S}\approx 50~$V. If we assume a conservative bias voltage $V_g=$5~V, we obtain 
\begin{equation}
N_\textrm{drive} = \frac{V_g}{2e}x_\textrm{zpf}\frac{dC}{dx}\simeq 0.01,
\end{equation}
where $x_\textrm{zpf}=\sqrt{\hbar/2m\Omega_m}$ is the mechanical mode zero point motion amplitude, and $\frac{dC}{dx}\simeq C/h$.

\bibliography{biblio.bib}

\end{document}